\documentclass[vecphys]{svmult}

\usepackage{makeidx}         
\usepackage{graphicx}        
\usepackage{multicol}        
\usepackage{cite}            
\usepackage[bottom]{footmisc}
\usepackage{url}
\usepackage[clockwise]{rotating}

\makeindex             

\begin{document}

\title{Neutron Scattering and Its Application to Strongly Correlated Systems}
\titlerunning{Neutron Scattering}
\toctitle{Neutron Scattering}

\author{Igor A. Zaliznyak \and John M. Tranquada}
\institute{Brookhaven National Laboratory, Upton, NY 11973-5000, USA \\
\texttt{zaliznyak@bnl.gov} (631-344-3761),
\texttt{jtran@bnl.gov} (631-344-7547)}

\maketitle

\begin{abstract}
Neutron scattering is a powerful probe of strongly correlated systems.  It can directly detect common phenomena such as magnetic order, and can be used to determine the coupling between magnetic moments through measurements of the spin-wave dispersions.  In the absence of magnetic order, one can detect diffuse scattering and dynamic correlations.  Neutrons are also sensitive to the arrangement of atoms in a solid (crystal structure) and lattice dynamics (phonons).  In this chapter, we provide an introduction to neutrons and neutron sources.  The neutron scattering cross section is described and formulas are given for nuclear diffraction, phonon scattering, magnetic diffraction, and magnon scattering.  As an experimental example, we describe measurements of antiferromagnetic order, spin dynamics, and their evolution in the La$_{2-x}$Ba$_x$CuO$_4$ family of high-temperature superconductors.
\end{abstract}

\noindent

\section{Introduction}
\label{Tranq:sec:1}
A common symptom of correlated-electron systems is magnetism, and neutron scattering is the premiere technique for measuring magnetic correlations in solids.  With a spin angular momentum of $\frac12\hbar$, the neutron interacts directly with the magnetization density of the solid.  Elastic scattering can directly reveal static magnetic order; for example, neutron diffraction provided the first experimental evidence for N\'eel antiferromagnetism \cite{Tranq:shul49}.  Through inelastic scattering one can probe dynamic spin-spin correlations; in an ordered antiferromagnet, one can measure the precession of the spins about their average orientations, which show up as dispersing spin waves.

Neutrons do not couple to the charge of the electrons, but instead scatter from atomic nuclei via the strong force.  Despite the name, the small size of the nucleus compared to the electronic charge cloud of the atom results in a rather weak scattering cross section.  The magnetic and nuclear scattering cross sections are comparable, so that neutron scattering is very sensitive to magnetism, in a relative sense.

A challenge with neutron scattering is that the combination of weak scattering cross section and limited source strength means that one needs a relatively large sample size compared with many other techniques.   The value of the information that can be obtained by neutron scattering generally makes worthwhile the effort to grow large samples; nevertheless, in practice it is useful to take advantage of complementary information obtained from techniques such as muon spin rotation spectroscopy and nuclear magnetic resonance.  The latter techniques yield less information but often provide greater precision.  There have also been continuing developments in resonant x-ray scattering; nevertheless, neutron scattering will remain an essential technique to investigate strongly correlated systems for the foreseeable future.

As we have space only for a concise introduction to the field, we note that there plenty of more extended references available.  A good summary of the theory of neutron scattering is given by Squires \cite{Tranq:squi12}, while a more detailed description is provided by Lovesey  \cite{Tranq:love84}.  We have contributed to a technique-oriented book \cite{Tranq:shir02} and to book chapters on magnetic neutron scattering \cite{Tranq:zali05,Tranq:zali07}, and new books on the subject continue to appear.

To illustrate some of the concepts and capabilities, we will use examples involving copper-oxide compounds, especially from the family La$_{2-x}$Ba$_x$CuO$_4$, which includes phenomena from antiferromagnetic order to high-temperature superconductivity.  More details on neutron scattering studies of cuprates are given in recent reviews \cite{Tranq:birg06,Tranq:tran07,Tranq:fuji12a,Tranq:bour11}.

\section{Basic properties of the neutron and its interaction with matter}
\label{Tranq:sec:2}\index{neutron properties}

\begin{table}[!pb]
\centering
\caption[]{Basic properties of a neutron. The gyromagnetic ratio, $\gamma_n$, and the g-factor, $g_n$, are defined by $\vec{\mu}_n = \gamma_n \vec{\sigma}_n = -g_n \mu_N {\bf S}_n $, where $\vec{\sigma}_n$ is the neutron's angular momentum, ${\bf S}_n = \vec{\sigma}_n /\hbar$ is the neutron's spin ($S_n = 1/2$), $\mu_N  = e\hbar/(2m_pc) = 5.05078 \times 10^{-27}$ J/T $= 5.05078 \times 10^{-24}$ erg/Gs is the nuclear magneton \cite{Tranq:mohr12,Tranq:wiet11}.}
\renewcommand{\arraystretch}{1.2}
\setlength\tabcolsep{5pt}
\begin{tabular}{@{}cccccc@{}}
\hline\noalign{\smallskip}
Charge & Mass & Lifetime & Magnetic & Gyromagnetic & g-factor \\
              &            &                 & moment $\mu_n$  & ratio $\gamma_n$ & $g_n$ \\
  & (kg) & (s) &  (J/T) &   (s$^{-1}$/T) &  \\
\hline\noalign{\smallskip}
0 & $1.67492 \times 10^{-27}$  & $882 \pm 2$ & $-0.96623 \times 10^{-26}$ & $-1.83247 \times 10^{8}$ &  3.82609 \\
\hline
\end{tabular}
\label{Tranq:tab:2-1}       
\end{table}

The neutron is an elementary spin-1/2 particle, which, together with its charged relative, the proton, is a building block of the atomic nucleus. According to the ``standard model'' of the elementary particles, the neutron and proton are fermionic hadrons, or baryons, composed of one ``up'' and two``down'' quarks, and two ``up'' and one ``down'' quarks, respectively. The basic properties of a neutron are summarized in Table~\ref{Tranq:tab:2-1}.

Although the neutron is electrically neutral, it has a non-zero magnetic moment, similar in magnitude to that of a proton ($\mu_n = 0.684979 \mu_p$), but directed opposite to the angular momentum, so that the neutron's gyromagnetic ratio is negative. The neutron's mass, $m_n = 1.00866$ Da (atomic mass units) is slightly larger than that of the proton, $m_n = 1.00728$ Da, and that of the hydrogen atom $m_{H} = 1.00782$ Da. Therefore, outside the nucleus the free neutron is unstable and undergoes $\beta-$decay into a proton, an electron, and an antineutrino. Although the free neutron's lifetime is only about 15 minutes, this is long enough for neutron-scattering experiments. For example, a neutron extracted through the beam-tube in a nuclear reactor has typically reached thermal equilibrium with the water that cools the reactor in a number of collisions on its way out (such neutrons usually are called thermal neutrons). Assuming the water has ``standard'' temperature of 293 K, the neutron's most probable velocity would be about 2200 m/s. It would spend only a fraction of a second while it travels along the $< 100$ m beam path in the spectrometer to be scattered by the sample and arrive in the detector.

Neutrons used in scattering experiments are non-relativistic. Therefore, the neutron's energy, $E_n$, is related to its velocity, $\vec{v}_n$, wave vector, $\vec{k}_n=m_n \vec{v}_n/\hbar$, and the (de Broglie) wavelength, $\lambda_n=2\pi/k_n$, through
\begin{equation}%
E_n = \frac{1}{2} m_n \vec{v}_n^2 = \frac{\hbar^2 \vec{k}_n^2}{2 m_n} = \frac{h^2}{2 m_n \lambda_n^2} \ts.
\label{Tranq:Eq:2-1}
\end{equation}%
Following the notation accepted in particle physics, the neutron's energy is measured in millielectronvolts (meV). The neutron's wavelength and its wave vector are usually measured in \AA\ (1~\AA$\null = 0.1$~nm $= 10^{-8}$ cm) and \AA$^{-1}$, respectively. Using these units, we can rewrite the Eq.~(\ref{Tranq:Eq:2-1}) in the following, practical fashion:
\begin{equation}%
E_n = 5.22704 \cdot 10^{-6} \cdot v_n^2 = 2.07212 \cdot k_n^2 = \frac{81.8042}{\lambda_n^2} \ts,
\label{Tranq:Eq:2-2}
\end{equation}%
where $E_n$ is in meV, $v_n$ in m/s, $k_n$ in \AA$^{-1}$, and $\lambda_n$ in \AA.
\index{neutron energy formulas}

For the sake of comparison with the notations used in other techniques and in theoretical calculations, we list several different ways of representing typical neutron energies in Table \ref{Tranq:tab:2-2}. The different energy equivalents shown in the Table can be used interchangeably, as a matter of convenience.

\begin{table}[!pt]
\centering
\caption[]{Different notations used to represent the neutron's energy. $e$ is the electron charge, $h$ is the Plank's constant, $c$ is the velocity of light, $\mu_B  = {e^2}/{2m_e c} = 0.92740 \times 10^{-29}$ J/T is the Bohr's magneton, $k_{\rm B}$ is Boltzman's constant \cite{Tranq:mohr12}. Also shown are the corresponding neutron wave vector $k_n$ and the deBroglie wavelength $\lambda_n$.}
\renewcommand{\arraystretch}{1.2}
\setlength\tabcolsep{5pt}
\begin{tabular}{@{}cccccccc@{}}
\hline\noalign{\smallskip}
 $E_n$ & $E_n/e$ & $E_n/h$ & $E_n/(hc)$ & $E_n/(2\mu_B)$ & $E_n/k_B$ & $k_n$ & $\lambda_n$ \\
 ($10^{-19}$ J) & (meV) & (THz) & (cm$^{-1}$) & (T) & (K) & (\AA$^{-1}$) & (\AA) \\
\hline\noalign{\smallskip}
 1.60218 & 1000 & 241.799 & 8065.54 & 8637.99 & 11604.5 & 21.968 & 0.2860 \\
 0.160218 & 100 & 24.1799 & 806.554 & 863.799 & 1160.45 & 6.9469 & 0.9044 \\
 0.0801088 & 50 & 12.0899 & 403.277 & 431.900 & 580.225 & 4.9122 & 1.27909 \\
 0.0240326 & 15 & 3.62698 & 120.983 & 129.570 & 174.068 & 2.6905 & 2.3353 \\
 0.00160218 & 1 & 0.241799 & 8.06554 & 8.63799 & 11.6045 & 0.69469 & 9.0445 \\
\hline
\end{tabular}
\label{Tranq:tab:2-2}  \index{neutron energy units}     \end{table}

\section{Neutron sources}
\index{neutron sources}

Neutrons are especially abundant in nuclei of high atomic number, where they can significantly exceed the number of protons.  To create a neutron beam, the first challenge is to extract neutrons from the nuclei.  The first practical source was the nuclear reactor, in which neutron bombardment of $^{235}$U nuclei induces fission, a process that releases several neutrons per incident neutron, thus allowing for a self-sustaining chain reaction.  The neutrons that are released have a very large energy, whereas the fission cross section is enhanced by slower neutrons.  The slowing of neutrons can be achieved quite effectively by scattering from hydrogen, especially in the form of H$_2$O, which can also act to cool the reactor core.  In a research reactor, where one would like to extract some of the neutrons, the reactor moderator can be made more transparent to neutrons by replacing H$_2$O with D$_2$O (heavy water, with D representing deuterium).   Cylindrical thimbles poking into the water moderator provide an escape path for neutrons, which form the beams that supply neutron spectrometers.  

Another approach is to knock the neutrons out of heavy nuclei with high-energy protons from an accelerator.  Again, the neutrons that can escape the nuclei have very high energies that must be reduced by multiple scattering in a moderator.  In contrast to a reactor, which produces neutron beams that are continuous in time, the proton beam provided by an accelerator can be pulsed, so that a spallation source typically has pulsed beams of neutrons.   Targets can be made of a heavy metal such as tungsten, but newer sources with higher power tend to use liquid mercury in order to allow adequate heat removal.

A list of the major operating spallation sources in the world is given in the upper portion of Table~\ref{Tranq:tb:source}.  Information on the available instrumentation and capabilities can be obtained from the listed web sites.  With a pulsed neutron source, each burst of neutrons is produced in a narrow time window, so that one can distinguish between neutrons of different velocities by their travel time, or ``time of flight''.  Using a rotating shutter, one can select incident neutrons of a desired energy; the energy of scattered neutrons can then be determined by their time of arrival at a detector.

\begin{sidewaystable}
\null\vskip 12cm
\centering
\caption[]{Major neutron user facilities presently in operation.}
\renewcommand{\arraystretch}{1.2}
\setlength\tabcolsep{5pt}
\begin{tabular}{@{}llll@{}}
\hline\noalign{\smallskip}
Facility & Laboratory & Location & Web address  \\
\hline\noalign{\medskip}
\smallskip \hskip 5 mm {\it Spallation Sources}\\
Spallation Neutron Source & Oak Ridge National Lab & Oak Ridge, TN, USA & http://neutrons.ornl.gov/ \\
Lujan Neutron Scattering Center & Los Alamos National Lab & Los Alamos, NM, USA & http://lansce.lanl.gov/lujan/ \\
ISIS & Rutherford Appleton Lab & Didcot, UK & http://www.isis.stfc.ac.uk/  \\
J-PARC & Japan Atomic Energy Agency  & Tokai, Japan & http://j-parc.jp/MatLife/en/ \\
SINQ & Paul Scherrer Institut & Villigen, Switzerland & http://www.psi.ch/sinq/ \\
\null\\ \smallskip
\hskip 5 mm {\it Reactor Facilities}\\
High Flux Isotope Reactor & Oak Ridge National Lab & Oak Ridge, TN, USA & http://neutrons.ornl.gov/ \\
NIST Center for Neutron Research & NIST & Gaithersburg, MD, USA & http://www.ncnr.nist.gov/ \\
Institut Laue Langevin & ILL & Grenoble, France & http://www.ill.eu/ \\
FRM-II & Technische Universit\"at   & Munich, Germany & http://www.frm2.tum.de/en/ \\
 & \quad\quad M\"unchen & & \\
Laboratoire L\'eon Brillouin & CEA Saclay & Saclay, France & http://www-llb.cea.fr/en/ \\
JRR-3 & Japan Atomic Energy  & Tokai, Japan & http://qubs.jaea.go.jp/en\_index.html \\
 & \quad\quad Agency & & \\
OPAL & ANSTO & Lucas Heights, NSW,  & http://www.ansto.gov.au/ \\
 & & \quad\quad Australia & \\
HANARO &KAERI & Daejeon, South Korea & http://hanaro4u.kaeri.re.kr/ \\
\hline
\end{tabular}
\label{Tranq:tb:source}       
\end{sidewaystable}

The spallation source SINQ at the Paul Scherrer Institut provides a continuous, rather than pulsed, beam, so its instrumentation has more in common with reactor facilities, which are listed in the lower portion of Table~\ref{Tranq:tb:source}.  With a continuous source, it is common to select the desired energy of incident neutrons by Bragg diffraction from a crystal (or array of crystals).  In a triple-axis spectrometer \cite{Tranq:shir02}, one also uses Bragg diffraction to analyze the energy and momentum of neutrons scattered by a sample.  Again, many of the facility web sites provide a wealth of information on spectrometers and capabilities.

\section{Neutron interactions and scattering lengths}

Many of the fundamental advantages of neutron scattering techniques arise from the fact that the neutron's interactions with matter are usually weak and are extremely well understood. Hence, neutrons afford direct experimental insight into dynamical properties of the material system of interest, unperturbed by the probe and essentially undistorted by the details of its interaction with matter. These properties contrast favorably with X-ray or charged-particle (electron, muon) techniques, where the probe could significantly perturb the system, and the interaction matrix elements between the system and the probe are often very complicated and profoundly impact the physics measured in the experiment.

The scattering of neutrons by an atomic system is governed by two fundamental interactions. The residual strong interaction (nuclear force) gives rise to scattering by the atomic nuclei (nuclear scattering). The electromagnetic interaction of the neutron's magnetic moment with the sample's internal magnetic fields, mainly originating from the unpaired electrons in the atomic shells, gives rise to magnetic scattering \cite{Tranq:sear89,Tranq:squi12,Tranq:izyu70,Tranq:love84,Tranq:jens91}.

\index{neutron magnetic interaction}

Magnetic interaction of a neutron with a single atom is of relativistic origin and is very weak, so that magnetic neutron scattering can be treated using the Born approximation. The interaction potential consists of the dipole-dipole interaction with the magnetic moment associated with the electronic spin, $\vec{\mu}_{se} = g_s \vec{s}_e \approx -2 \vec{s}_e$ ($g_s \approx -2.002319$ is the Land\'e $g$-factor),
\begin{equation}%
\hat{V}_{se} (\vec{r}) = -\frac{8\pi}{3} (\vec{\mu}_n \cdot \vec{\mu}_{se}) \delta(\vec{r}) - \frac{(\vec{\mu}_n \cdot \vec{\mu}_{se})}{r^3} +
 \frac{3(\vec{\mu}_n\cdot \vec{r}) (\vec{\mu}_{se}\cdot \vec{r})}{r^5} \ts,
\label{Tranq:Eq:V-spin}
\end{equation}%
and the interaction with the electric current associated with the electron's orbital motion
\begin{equation}%
\hat{V}_{sl} (\vec{r}) = 2\mu_B \frac{(\vec{\mu}_n \cdot \vec{l}_{e})}{r^3} \ts.
\label{Tranq:Eq:V-orb}
\end{equation}%
Here $\hbar \vec{l}_e = \vec{r} \times \vec{p}_e$ is the electron's orbital angular momentum, and $\vec{r} = \vec{r}_e - \vec{r}_n$ its coordinate in the neutron's rest frame.

\index{neutron scattering length}

While the neutron's interaction with the atomic nucleus is strong---the nuclear force is responsible for holding together protons and neutrons in the nucleus---it has extremely short range, $< 10^{-12}$ cm, comparable to the size of the nuclei, and is much smaller than the typical neutron's wavelength. Hence, to describe the neutron's interaction with the system of atomic nuclei in which the typical distances are about 1~\AA $= 10^{-10}$~cm, a highly accurate approximation is obtained by using a delta-function for the nuclear scattering length operator in the coordinate representation,
\begin{equation}%
\hat{b}_{N} (\vec{r}) = b\, \delta(\vec{r}_n - \vec{R}_N) \ts.
\label{Tranq:Eq:2-5}
\end{equation}%
Here $\vec{R}_N$ is the position of the nucleus and $b$ is the nuclear scattering length, which is usually treated as a phenomenological parameter \cite{Tranq:ferm47a,Tranq:ferm47b} that has been determined experimentally and tabulated \cite{Tranq:inte95,Tranq:sear92,Tranq:mugh06}. In the Born approximation, the scattering length would correspond to the neutron-nucleus interaction described by the Fermi's {\it pseudo-potential} \cite{Tranq:ferm36},
\begin{equation}%
\hat{V}_{N} (\vec{r}_n, \vec{R}_N) = - \frac{2\pi \hbar^2}{m_n} b\, \delta(\vec{r}_n - \vec{R}_N) \ts.
\label{Tranq:Eq:2-6}
\end{equation}%
In general, the bound scattering length (that is, for a nucleus fixed in space) is a complex quantity \cite{Tranq:sear89,Tranq:squi12}, $b = b' - i b''$, defining the total scattering cross-section, $\sigma_s$, and the absorption cross-section far from the nuclear resonance capture, $\sigma_a$, through
\begin{equation}%
\sigma_s = 4 \pi |b'|^2 \;\;\; \sigma_a = \frac{4\pi}{k_i} |b''|^2 \ts.
\label{Tranq:Eq:2-7}
\end{equation}%
For the majority of natural elements $b'$ is close in magnitude to the characteristic magnetic scattering length, $r_m = -(g_n/2) r_e = - 5.391$ fm (1 fm $= 10^{-13}$ cm and $r_e=e^2/(m_e c^2)$ is the classical electron radius).

\index{neutron scattering length}
\index{neutron scattering cross section}
\index{neutron absorption cross section}

\section{Cross-section measured in a neutron scattering experiment}

\begin{figure}[t]
\centering
\includegraphics*[width=1.\textwidth]{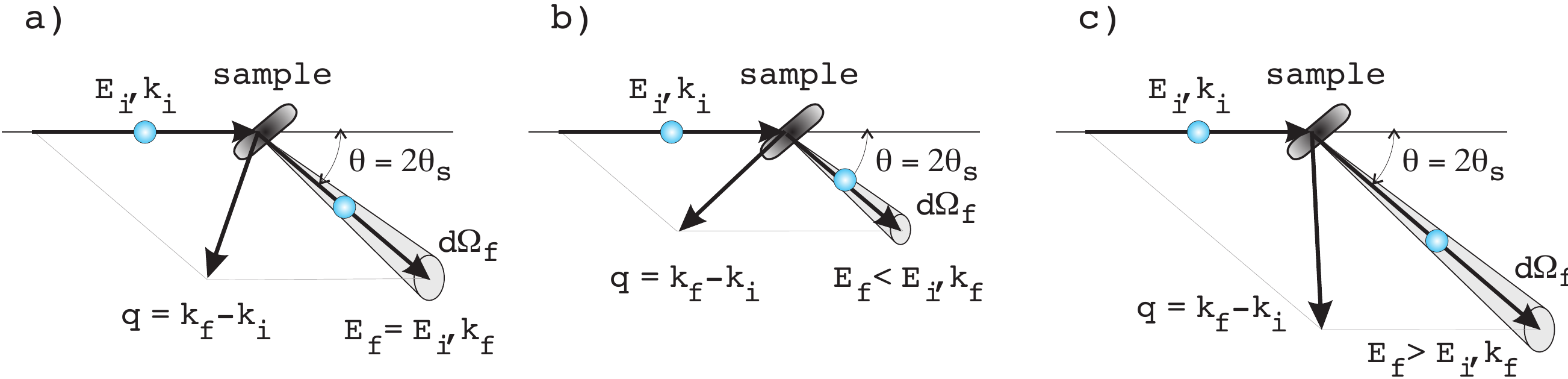}
\caption[]{Schematics of the scattering process in a neutron scattering experiment, (a) elastic, (b), inelastic, neutron energy loss, (c), inelastic, neutron energy gain.}
\label{Tranq:Fig1:scattering}
\end{figure}

In a scattering experiment, the sample is placed in the neutron beam having a well-defined wave vector $\vec{k}_i$ and known incident flux density $\Phi_i(\vec{k}_i)$, and the detector measures the partial current, $\delta J_f(\vec{k}_f)$, scattered into a small (ideally infinitesimal) volume of the phase space, $d^3\vec{k}_f = k_f^2 dk_f d\Omega_f = (m_n k_f/\hbar^2) dE_f d\Omega_f$, near the wave vector $\vec{k}_f$, as indicated in Fig. \ref{Tranq:Fig1:scattering}. 
This measured partial current, normalized to the appropriate phase space element covered by the detector, yields the scattered current density. The double differential scattering cross-section, which is thus measured, is then defined by the ratio of this scattered current density to the incident neutron flux density, e.g.,
\begin{equation}%
\frac{d^2 \sigma (\vec{Q}, E)}{dE d\Omega} = \frac{1}{\Phi_i(\vec{k}_i)} \frac{\delta J_f(\vec{k}_f)}{dE d\Omega} \ts.
\label{Tranq:Eq:CrossSec-1}\index{neutron scattering, differential cross section}
\end{equation}%
For each incident neutron in the plane wave state $e^{i \vec{k}_i\cdot\vec{r}_n}$, the incident flux density is $\Phi_i(\vec{k}_i) = \hbar k_i/m_n$. The scattered current density is determined by the transition rate $\Gamma_{i \rightarrow f}$ from the initial state $|\vec{k}_i, S^z_{n,i}, \eta_i \rangle$, where the neutron is in the plane wave state $e^{i \vec{k}_i\cdot \vec{r}_n}$ with the spin $S^z_{n,i}$ and the scattering system is described by the set of variables ${\eta_i}$, to the final state, $|\vec{k}_f, S^z_{n,f}, \eta_f \rangle$. According to scattering theory \cite{Tranq:sear89,Tranq:lipp50,Tranq:zali05}, the transition rate is determined by the matrix elements of the transition operator (or $T$-matrix) $\hat{T}$, satisfying certain operator equations, which depend on the scattering system's Hamiltonian, $\hat{H}$, and its interaction with the neutron, $\hat{V}$,
\begin{equation}%
\Gamma_{i \rightarrow f} =
 \frac{2\pi}{\hbar}\left|\langle \vec{k}_f, S^z_{n,f}, \eta_f | \hat{T} |\vec{k}_i, S^z_{n,i}, \eta_i \rangle \right|^2
 \delta \left( \frac{\hbar^2 k_i^2}{2m_n} - \frac{\hbar^2 k_f^2}{2m_n} - E \right) \ts.
\label{Tranq:Eq:TransitionRate}
\end{equation}%
Here $E = E_f(\eta_f) - E_i(\eta_i)$ is the scattering system's energy gain. It is convenient to introduce the scattering length operator, $\hat{b}$, which conveniently absorbs several factors, 
\begin{equation}%
\hat{b}(\vec{r}_n, \vec{S}_n, \eta) = - \frac{m_n}{2 \pi \hbar^2} \langle \vec{k}_f, S_{n,f}^z | \hat{T} |\vec{k}_i,S^z_{n,i}\rangle \ts, 
\end{equation}
and its Fourier transform, $\hat{b}(\vec{q})$,
\begin{equation}
\hat{b}(\vec{q}) = \int {\rm e}^{-i\vec{q}\cdot\vec{r}} \hat{b} (\vec{r}, \vec{S}_n, \eta) d^3 \vec{r}
\ts,
\label{Tranq:Eq:ScatLength}
\end{equation}%
Summing over all possible final scattering states, we obtain the double differential scattering cross-section for a given initial state, $|\vec{k}_i, S^z_{n,i}, \eta_i \rangle$,
\begin{equation}%
\frac{d^2 \sigma (\vec{Q}, E)}{dE d\Omega} = \frac{k_f}{k_i}\sum_{S_{n,f}^z,\eta_f}  \left|\langle \eta_f | \hat{b}(-\vec{Q}) | \eta_i \rangle \right|^2
 \delta \left( E_f(\eta_f) - E_i(\eta_i) - E \right) \ts,
\label{Tranq:Eq:CrossSec-2}\index{neutron scattering, differential cross section}
\end{equation}%
where the dependence on the spin-state of the neutron is implicit in $\hat{b}(-\vec{Q})$.
The energy and momentum transfer to the sample are governed by the conservation laws,
\begin{equation}%
\vec{Q} = \vec{k}_i - \vec{k}_f \ts,\;\;\; E = E_f(\eta_f) - E_i(\eta_i) = \frac{\hbar^2 }{2m_n}(k_i^2 - k_f^2) \ts.
\label{Tranq:Eq:QE-conservation}
\end{equation}%
Finally, following Van Hove \cite{Tranq:vanh54}, one can use the integral representation of the delta-function expressing the energy conservation in Eq.~(\ref{Tranq:Eq:CrossSec-2}), and the time-dependent scattering length operator whose evolution is governed by the system's Hamiltonian,
\begin{equation}%
\hat{b}(\vec{q}, t) = {\rm e}^{ i \hat{H} t/\hbar} \hat{b}(\vec{q}) {\rm e}^{ - i\hat{H} t/\hbar} \ts,
\label{Tranq:Eq:Heisenberg-b}
\end{equation}%
to recast the double differential scattering cross-section in the most useful form of the two-time correlation function,
\begin{equation}%
\frac{d^2 \sigma }{dE d\Omega} =  \frac{k_f}{k_i} \sum_{S_{n,f}^z} \int_{-\infty}^{\infty} {\rm e}^{ -i\omega t} \langle  \eta_i | \hat{b}^\dag (-\vec{Q})  \hat{b}(-\vec{Q}, t) |\eta_i \rangle \frac{dt}{2 \pi \hbar} \ts.
\label{Tranq:Eq:CrossSec-3}\index{neutron scattering, differential cross section}
\end{equation}%
Here the sum is over all possible final spin states of the scattered neutron, $S^z_{n,f}$, since in the general case the scattering length operator, $\hat{b}(-\vec{Q}, t)$, depends on the neutron spin, $\vec{S}_{n}$. The sum over the final states of the sample has been absorbed into the expectation value of the two-time correlation function of the scattering length operator. The minus sign in front of $\vec{Q}$ in Eq.~(\ref{Tranq:Eq:CrossSec-3}) follows from the convention adopted in the conservation laws in Eq.~(\ref{Tranq:Eq:QE-conservation}), where $\hbar \vec{Q}$ is the momentum transfer to the sample, which is the opposite of the change in the neutron's momentum. The total measured scattering cross-section is obtained by taking the proper thermal average of Eq.~(\ref{Tranq:Eq:CrossSec-3}) over all possible initial states, $|\eta_i \rangle$.
 
While the scattered neutron's wave vector $\vec{k}_f$ is uniquely determined by $\vec{k}_i$ and $\vec{Q}$, by virtue of the conservation laws (\ref{Tranq:Eq:QE-conservation}), the neutron's spin state can be changed by transferring the angular momentum $\hbar (\Delta S^z_{n}) = \pm \hbar$ to the sample. In a polarized neutron experiment scattering between different neutron spin states can be measured. In such a case, the scattering length operator in Eq.~(\ref{Tranq:Eq:CrossSec-3}) is a matrix with respect to different initial and final spin state indices; it determines the various spin-flip and non-spin-flip cross-sections \cite{Tranq:love84,Tranq:blum61}. In the more common case of unpolarized neutron scattering, neutron spin indices should be traced out in Eq.~(\ref{Tranq:Eq:CrossSec-3}), so that it determines a single unpolarized neutron scattering cross-section.
 
Finally, we should mention that the double differential cross-sections in Eq.~(\ref{Tranq:Eq:CrossSec-2}), (\ref{Tranq:Eq:CrossSec-3}) are general expressions obtained from scattering theory and are valid for scattering of any probe particles. The remarkable advantage of neutron scattering is in the fact that scattering length operators are rather simple, very well understood, and are directly related to the fundamental physical properties of the scattering sample.

\section{Nuclear scattering in condensed matter}
\index{neutron nuclear scattering}

\begin{figure}[t]
\centering
\includegraphics*[width=0.75\textwidth]{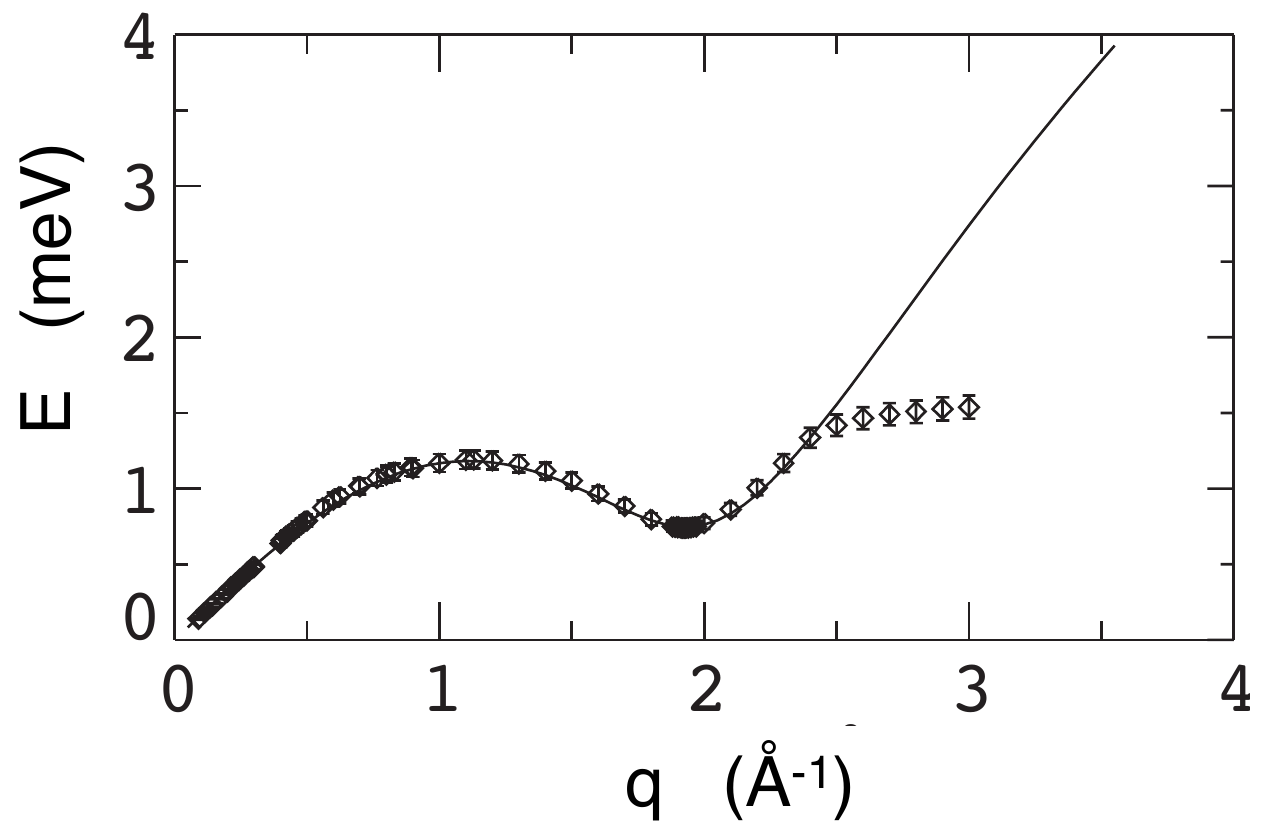}
\caption[]{Phonon-roton dispersion of the elementary excitations in the superfluid $^4$He. The points show the compilation of the experimental neutron data presented in Ref. \cite{Tranq:donn81}. The solid line is the fit of the low-$q$ part of the spectrum to the Bogolyubov quasiparticle dispersion.}
\label{Tranq:Fig:HeliumDisp}\index{superfluid helium, phonon-roton dispersion}
\end{figure}

For scattering from an individual nucleus, the scattering length operator can be very accurately approximated by a delta-function, Eq.~(\ref{Tranq:Eq:2-5}). For a collection of nuclei in a condensed matter system, the total scattering length operator is obtained by adding scattering lengths of all nuclei,
\begin{equation}%
\hat{b}_{N} (\vec{r}_n) = \sum_{j} b_{j}\, \delta(\vec{r}_n - \vec{r}_{j}) \ts,
\label{Tranq:Eq:b-nuclei}
\end{equation}%
where $j$ indexes the nucleus at position $\vec{r}_{j}$ with scattering length $ b_{j}$. For a system of identical nuclei, this is just a particle number density operator in the scattering system, times $b$,
\begin{equation}%
\hat{b}_{N} (\vec{r}_n) = b\, \sum_{j} \delta(\vec{r}_n - \vec{r}_{j}) =  b\, \hat{n}(\vec{r}_n) \ts, \;\;\;
\hat{b}_{N} (\vec{q}) =  b\, \hat{n}_{\vec{q}} \ts.
\label{Tranq:Eq:b-nuclei-n}
\end{equation}%
Substituting this into Eq.~(\ref{Tranq:Eq:CrossSec-3}) and summing out the neutron's spin states we obtain,
\begin{equation}%
\frac{d^2 \sigma }{dE d\Omega} = \frac{k_f}{k_i} |b|^2 \int_{-\infty}^{\infty} {\rm e}^{ -i\omega t} \langle \eta_i | \hat{n}_{\vec{Q}} \hat{n}_{-\vec{Q}}(t) | \eta_i \rangle \frac{dt}{2 \pi \hbar} \ts.
\label{Tranq:Eq:CrossSec-n}
\end{equation}%
Therefore, the nuclear cross-section measures the space-time correlation of the atom number density in a condensed matter system. This is exactly the quantity of interest in many theories of strongly-correlated quantum systems. 

One of the first successes of neutron scattering was the measurement of the phonon-roton dispersion of the elementary excitations in superfluid helium-4. Neutron data have confirmed that the shape of the dispersion is in agreement with that previously postulated by Landau and Feynman, as illustrated in Fig. \ref{Tranq:Fig:HeliumDisp}.  This led to the broad acceptance of the neutron scattering technique as a prime tool for studying quantum systems.

Next we consider the case in which two or more types of nuclear scatterers (with distinct scattering lengths $b_j$ and frequency of occurrence $c_j$) are present in the sample in a random fashion.  For example, an element may have multiple isotopes, each with a distinct $b_{j}$, or the nuclei have a spin, so that $b_{j}$ depends on the nuclear and neutron spin orientations, or we have at least two elements that are randomly distributed among equivalent positions.    The average product of the scattering lengths for any two sites can then be written as
\begin{equation}%
\overline{(b_{j} b_{j'})} = (\overline{b})^2 \, \left( 1 - \delta_{jj'} \right) +  \overline{b^2} \, \delta_{jj'} \ts,
\label{Tranq:Eq:b-nuclei-ave}
\end{equation}%
where
\begin{eqnarray}
\overline{b} & = &  \sum_{j} c_j b_j \ts, \nonumber \\
\overline{b^2} & = & \sum_{j} c_j b_j^2 \ts.
\end{eqnarray}
We can then distinguish between coherent scattering,
\begin{equation}%
\frac{d^2 \sigma_{c}}{dE d\Omega} = \frac{k_f}{k_i} (\overline{b})^2 \sum_{jj'} \int_{-\infty}^{\infty} {\rm e}^{ - i\omega t } \langle {\rm e}^{-i \vec{Q}\cdot\vec{r}_{j}} {\rm e}^{ i \vec{Q}\cdot\vec{r}_{j'}(t)} \rangle \frac{dt}{2 \pi \hbar} \ts,
\label{Tranq:Eq:CrossSec-n-coh}\index{neutron nuclear scattering, coherent}
\end{equation}%
which probes the inter-nuclear correlation, and the incoherent scattering,
\begin{equation}%
\frac{d^2 \sigma_{i} }{dE d\Omega} = \frac{k_f}{k_i} \left( \overline{b^2} - (\overline{b})^2 \right) \sum_{j} \int_{-\infty}^{\infty} {\rm e}^{ -i\omega t } \langle {\rm e}^{-i \vec{Q}\cdot\vec{r}_{j}} {\rm e}^{i \vec{Q}\cdot\vec{r}_{j}(t)} \rangle \frac{dt}{2 \pi \hbar} \ts,
\label{Tranq:Eq:CrossSec-n-incoh}\index{neutron nuclear scattering, incoherent}
\end{equation}%
which probes the local autocorrelation of the nuclear position; the angle brackets denote the average over the sample state. In Eqs.~(\ref{Tranq:Eq:CrossSec-n-coh}) and (\ref{Tranq:Eq:CrossSec-n-incoh}) we have switched to the co-ordinate representation of nuclear density operator (\ref{Tranq:Eq:b-nuclei-n}) and performed the Fourier integration. As a result, nuclear positions $\vec{r}_{j}$ and $\vec{r}_{j'}(t)$ are quantum-mechanical operators and have to be treated appropriately in calculating the cross-section \cite{Tranq:squi12,Tranq:izyu70,Tranq:love84}.

\section{Nuclear scattering in a crystal: the Bragg peaks and the phonons}

In a crystal, the equilibrium positions of atomic nuclei are arranged on the sites of a lattice, so that the position of each individual nucleus $j$ can be represented as
\begin{equation}%
\vec{r}_{j} = \vec{R}_{j} +  \vec{u}_{j} \ts,
\label{Tranq:Eq:R-nuclei}
\end{equation}%
where $\vec{R}_{j}$ is the lattice site position, and $\vec{u}_{j}$ is a small displacement of the atomic nucleus from its equilibrium position at $\vec{R}_{j}$.

Substituting this into Eq.~(\ref{Tranq:Eq:CrossSec-n-coh}), one can show that the coherent nuclear cross-section of a monoatomic crystal is given by
\begin{equation}%
\frac{d^2 \sigma_{c}}{dE d\Omega} = \frac{k_f}{k_i} N (\overline{b})^2 {\rm e}^{- \langle (\vec{Q}\cdot \vec{u}_{0})^2 \rangle} \sum_{j} {\rm e}^{-i \vec{Q}\cdot\vec{R}_{j}} \int_{-\infty}^{\infty} {\rm e}^{ - i\omega t } {\rm e}^{ \langle (\vec{Q}\cdot\vec{u}_0) (\vec{Q}\cdot\vec{u}_{j}(t)) \rangle} \frac{dt}{2 \pi \hbar} \ts,
\label{Tranq:Eq:CrossSec-n-coh-cryst}
\end{equation}%
Here $\langle (\vec{Q}\cdot\vec{u}_{0})^2 \rangle$ is the time- or lattice-averaged square of the atomic displacement from equilibrium in the direction of $\vec{Q}$, and we have taken advantage of the fact that the correlation function in Eq.~(\ref{Tranq:Eq:CrossSec-n-coh}) depends only on relative coordinates, which allows one summation over the $N$ lattice sites to be completed.  The integral contains an exponentiated correlation function of atomic displacements.  It is useful to consider the series expansion of this term in powers of pair displacement correlations.

In zeroth order, the exponential factor is just 1, and one obtains the expression for the elastic Bragg scattering in a crystal,
\begin{equation}%
\frac{d^2 \sigma_{B} }{dE d\Omega} = N (\overline{b})^2 {\rm e}^{- 2W} \sum_{j} {\rm e}^{-i \vec{Q}\cdot\vec{R}_{j}} \delta(\hbar\omega) \ts,
\label{Tranq:Eq:CrossSec-n-Bragg-1}\index{Bragg diffraction, neutron nuclear}
\end{equation}%
where we used the conventional notation for the Debye-Waller factor,\index{Debye-Waller factor} with $W \equiv \frac12 \langle (\vec{Q}\cdot\vec{u}_{0})^2 \rangle$. Using the lattice Fourier representation, this can be recast in the common form
\begin{equation}%
\frac{d^2 \sigma_{B} }{dE d\Omega} = 
N V^* (\overline{b})^2 {\rm e}^{- 2W} \sum_{\vec{\tau}} \delta(\vec{Q} - \vec{\tau}) \, \delta (\hbar\omega) \ts,
\label{Tranq:Eq:CrossSec-n-Bragg-2}
\end{equation}%
where $V^* = (2\pi)^3/V_0$ is the reciprocal unit cell's volume ($V_0$ is the volume of the unit cell in real space), and $\vec{\tau}$ are the vectors of the reciprocal lattice.
In a non-Bravais crystal, where the unit cell contains several atoms, the sum in Eq.~(\ref{Tranq:Eq:CrossSec-n-coh-cryst}) has to be split into the intra-unit cell and the inter-unit cell parts, leading to
\begin{equation}%
\frac{d^2 \sigma_{B} }{dE d\Omega} = NV^* |F_{N}(\vec{Q})|^2 \sum_{\vec{\tau}} \delta(\vec{Q} - \vec{\tau}) \, \delta (\hbar\omega) \ts,
\label{Tranq:Eq:CrossSec-n-Bragg-3}
\end{equation}%
where the intra-unit cell summation yields the nuclear unit cell structure factor,
\begin{equation}%
F_{N}(\vec{Q}) = \sum_{\mu} {\rm e}^{- W_\mu} \overline{b}_{\mu} {\rm e}^{-i \vec{Q}\cdot\vec{r}_{\mu}} \ts,
\label{Tranq:Eq:cell-struc-fac}\index{structure factor, neutron nuclear}
\end{equation}%
and $\mu$ indexes atoms in the unit cell. For some reciprocal lattice points, $F_{N}(\vec{\tau})$ can be zero, which gives the Bragg peak extinction rules in a non-Bravais crystal.

Expanding the exponent in Eq.~(\ref{Tranq:Eq:CrossSec-n-coh-cryst}) to the first order, we obtain a contribution to the cross-section that is proportional to the correlation of displacements at two different sites. In cases where static disorder is present in the crystal, such as dislocations or lattice strain, the time-independent correlations of displacements between different sites give rise to elastic diffuse scattering.

Calculation of the time-dependent displacements of atomic nuclei from their equilibrium positions in the lattice is achieved by quantizing their vibrations in terms of quantum oscillators, called phonons. A phonon is a normal mode of atomic vibration, a coherent wave of atomic displacements in the crystal. We distinguish phonons with index $s$.  The polarization vector $\vec{e}_s$ (direction of atomic displacements) and the dependence of the energy on the wave vector, $\hbar\omega_{\vec{q}s}$ (dispersion), are determined by the the local inter-atomic potentials. The total number of such modes depends on the number of atoms in the unit cell of the crystal. Only three phonons, which are all acoustic, are present for the Bravais lattice, two transverse and one longitudinal. Taking the proper thermal average over the sample's equilibrium state, the contribution to the neutron scattering is given by
\begin{eqnarray}
\frac{d^2 \sigma_{ph} }{dE d\Omega} & = & \frac{k_f}{k_i} (\overline{b})^2 {\rm e}^{- 2W}\sum_s \frac{(\vec{Q}\cdot{\vec{e}_s)^2}}{2 M \omega_{\vec{q}s}} \times \nonumber\\
 & & \quad V^* \sum_{\vec{\tau}} \left[ \delta(\vec{Q} - \vec{q} - \vec{\tau}) \delta(\hbar\omega - \hbar\omega_{\vec{q}s})(n(\omega) + 1) \right. \nonumber\\
 & & \qquad\qquad\quad \left. + \, \delta(\vec{Q} + \vec{q} - \vec{\tau}) \delta(\hbar\omega + \hbar\omega_{\vec{q}s}) n(\omega) \right]
\ts,
\label{Tranq:Eq:CrossSec-phonon-Brav}\index{phonons, neutron scattering cross section}
\end{eqnarray}%
where $M$ is the mass of each nucleus.  The thermal factor 
\begin{equation}
n(\omega) = ({\rm e}^{\hbar\omega/k_{\rm B}T} -1)^{-1} \nonumber
\end{equation} 
is the Bose distribution function describing thermal population of the oscillator states for temperature $T$ of the sample. The first term arises from phonon creation and corresponds to the neutron energy loss, while the second term is from an annihilation of a phonon that has been thermally excited in the crystal and results in the neutron energy gain.

For a non-Bravais crystal lattice, there are also optic phonons, arising from the different intra-unit-cell vibrations. The total number of phonons is equal to $3 \nu$, the number of vibrational degrees of freedom of the $\nu$ atoms comprising the basis of the unit cell of the lattice. The contribution of each of these phonons to the neutron scattering cross-section is
\begin{eqnarray}
\frac{d^2 \sigma_{ph} }{dE d\Omega} & = & \frac{k_f}{k_i} \left| \sum_\mu \frac{\overline{b}_{\mu} \, {\rm e}^{- W_{\mu}} }{\sqrt{2 M_{\mu} \omega_{\vec{q}s}}} (\vec{Q}\cdot\vec{e}_{s\mu}) {\rm e}^{-i \vec{Q}\cdot \vec{r}_{\mu}} \right|^2 \times \nonumber\\
 & & \quad V^* \sum_{\vec{\tau}} \left[ \delta(\vec{Q} - \vec{q} - \vec{\tau}) \delta(\hbar\omega - \hbar\omega_{\vec{q}s})(n(\omega) + 1) \right. \nonumber\\
 & & \qquad\qquad\quad \left. + \, \delta(\vec{Q} + \vec{q} - \vec{\tau}) \delta(\hbar\omega + \hbar\omega_{\vec{q}s}) n(\omega) \right] \ts,
\label{Tranq:Eq:CrossSec-phonon-nonBrav}
\end{eqnarray}%
where $\vec{e}_{s\mu}$ is the polarization vector for site $\mu$ in mode $s$. For an acoustic phonon in the hydrodynamic, long-wavelength (small $q$) and low-energy limit, this reduces to Eq.~(\ref{Tranq:Eq:CrossSec-phonon-Brav}), where the total mass of all atoms in the unit cell, $M = \sum_\mu M_\mu$, should be used and one must multiply by the square of the elastic structure factor, $|F_N(\vec{Q})|^2$.

\section{Magnetic scattering in a crystal. Magnetic form factor and spin correlations.}
\index{neutron magnetic scattering}

The magnetic interaction of a neutron with a single atom is very weak, so the Born approximation provides an extremely accurate account for magnetic neutron scattering by the atomic electrons. In this approximation, the transition matrix is given simply by the interaction potential, $\hat{T} = \hat{V}$, where we have to combine the neutron's interaction with the electron's spin and orbital magnetic moment, Eqs. (\ref{Tranq:Eq:V-spin}) and (\ref{Tranq:Eq:V-orb}).  Accurate accounting for the orbital contribution to magnetic scattering presents, in general, a rather difficult and cumbersome task \cite{Tranq:love84}.  There are some important cases where the orbital contribution is not significant, such as transition-metal atoms in a crystal, where the local crystal electric field typically quenches the orbital angular momentum, or the case of $s$-electrons, where $l = 0$. Nevertheless, under some very general assumptions, the neutron's interaction with the electron orbital currents can be recast in the same way as its interaction with the spin magnetic moment, yielding for the total magnetic scattering length,
\begin{eqnarray}%
\hat{b}_{m} (\vec{r}) & = & - \frac{m_n}{2 \pi \hbar^2} \left( \hat{V}_{se} (\vec{r}) + \hat{V}_{le} (\vec{r}) \right) \nonumber \\
 & = & \frac{m_n}{2 \pi \hbar^2} \left( \vec{\mu}_n \cdot \sum_{e} \left[ \vec{\nabla} \times \left[ \vec{\nabla} \times \frac{\vec{\mu}_e (r)}{r} \right] \right] \right) \ts,
\label{Tranq:Eq:b-mag}\index{magnetic scattering length}
\end{eqnarray}%
where $\vec{\mu}_e (r) = \vec{\mu}_{s,e} + \vec{\mu}_{l,e} $ is the sum of the spin and the orbital magnetization associated with each electron, $e$. The Fourier transform of the magnetic scattering length (\ref{Tranq:Eq:b-mag}), which determines the scattering cross-section, is
\begin{equation}%
\hat{b}_{m} (\vec{Q}) = - \frac{m_n}{2 \pi \hbar^2} \frac{4 \pi}{Q^2} \left( \vec{\mu}_n \cdot \left[ \vec{Q} \times \left[ \vec{Q} \times \vec{m} (\vec{Q}) \right] \right] \right) \ts.
\label{Tranq:Eq:b-mag-q}
\end{equation}%
Here $\vec{m} (\vec{Q})$ is the Fourier transform of the total magnetization density of the atom,
\begin{eqnarray} 
\vec{m} (\vec{Q}) & =&  \vec{m}_S (\vec{Q}) + \vec{m}_L (\vec{Q}) \nonumber \\
 & = & \int {\rm e}^{-i\vec{Q}\cdot\vec{r}} \sum_e \left( -2\mu_B \vec{s}_e \delta( \vec{r} -  \vec{r}_e) + \vec{\mu}_{l,e} \right) d^3 \vec{r}, \nonumber
\end{eqnarray} 
$\vec{s}_e$ is the spin operator of $e-$th electron, $\vec{\mu}_{l,e}$ its orbital magnetic moment operator.

The cross product in Eq.~(\ref{Tranq:Eq:b-mag-q}) ensures the important property that only magnetization perpendicular to the wave vector transfer, $\vec{Q}$, contributes to the magnetic neutron scattering. Adding the contributions from all atoms in the crystal and averaging over the neutron polarizations, we obtain the magnetic neutron scattering cross-section measured in an experiment with unpolarized neutrons ($\alpha, \beta = x, y, z$),
\begin{equation}%
\frac{d^2 \sigma_{m} }{dE d\Omega} = \frac{k_f}{k_i} \left( \frac{2 m_n}{\hbar^2} \mu_n \right)^2 \sum_{\alpha,\beta} \left(\delta_{\alpha \beta} - \frac{Q_{\alpha} Q_{\beta}}{Q^2} \right) \int_{-\infty}^{\infty} {\rm e}^{ -i\omega t } \langle M^\alpha_{\vec{Q}} M^\beta_{-\vec{Q}} (t) \rangle \frac{dt}{2 \pi \hbar} \ts.
\label{Tranq:Eq:CrossSec-mag}\index{neutron magnetic scattering cross section}
\end{equation}%
Here 
\begin{eqnarray}
\vec{M}_{\vec{Q}} & = & \sum_{j} {\rm e}^{ -i \vec{Q}\cdot\vec{R}_j }\vec{m}_{j} (\vec{Q}) \nonumber \\
 & = & \int  {\rm e}^{ -i \vec{Q}\cdot\vec{r} } \sum_j \vec{m}_{j}(\vec{r} + \vec{R}_j) d^3 \vec{r} \nonumber
\end{eqnarray}
is the Fourier transformed magnetization density operator in the crystal. Hence, magnetic neutron scattering measures the time- and space-dependent correlations of the magnetization fluctuations in the sample. Introducing the dynamic correlation function,
\begin{equation}%
S^{\alpha\beta}(\vec{Q},\omega) = \int_{-\infty}^{\infty} {\rm e}^{-i\omega t} \langle M^\alpha_{\vec{Q}} M^\beta_{-\vec{Q}} (t) \rangle \frac{dt}{2 \pi \hbar} \ts,
\label{Tranq:Eq:SMM-corr}\index{dynamic correlation function}
\end{equation}%
we can rewrite Eq.~(\ref{Tranq:Eq:CrossSec-mag}) as
\begin{equation}%
\frac{d^2 \sigma_{m} }{dE d\Omega} = \frac{k_f}{k_i} r_m^2 \sum_{\alpha,\beta} \left(\delta_{\alpha \beta} - \frac{Q_{\alpha} Q_{\beta}}{Q^2} \right) \frac{1}{(2 \mu_B)^2} S^{\alpha\beta} (\vec{Q},\omega) \ts,
\label{Tranq:Eq:CrossSec-mag-S}
\end{equation}%
where $r_m = -2 \mu_B \mu_n ({2 m_n}/{\hbar^2}) = -5.391 \cdot 10^{-13}$ cm is the characteristic magnetic scattering length.

\subsection{The detailed balance constraint and the FDT}

The dynamic correlation function defined above by Eq.~(\ref{Tranq:Eq:SMM-corr}) obeys two important relations that are derived in the linear response theory \cite{Tranq:squi12,Tranq:love84,Tranq:jens91}. First, it is the detailed balance constraint, which relates the energy gain and the energy loss scattering at a temperature $T$,
\begin{equation}%
S^{\alpha\beta} (\vec{Q}, \omega) = {\rm e}^{{\hbar\omega}/{k_{\rm B}T}} S^{\beta\alpha}(-\vec{Q}, -\omega) \ts.
\label{Tranq:Eq:DelBal}\index{detailed balance}
\end{equation}%
The second is the fluctuation-dissipation theorem (FDT), which relates the scattering intensity with the imaginary part of the dynamic magnetic susceptibility,
\begin{equation}%
\tilde{\chi}^{''}_{\alpha\beta} (\vec{Q}, \omega) = \pi \left( 1 - {\rm e}^{-{\hbar\omega}/{k_{\rm B}T}} \right) \tilde{S}^{\alpha\beta}(\vec{Q}, \omega) \ts.
\label{Tranq:Eq:FDT}\index{fluctuation-dissipation theorem}\index{magnetic susceptibility, dynamic}
\end{equation}%
Here $\tilde{\chi}^{''}_{\alpha\beta} (\vec{Q}, \omega)$ and $\tilde{S}^{\alpha\beta}(\vec{Q}, \omega)$ denote ${\chi}^{''}_{\alpha\beta} (\vec{Q}, \omega)$ and ${S}^{\alpha\beta} (\vec{Q}, \omega)$ symmetrized with respect to $\{\alpha, \beta, \vec{Q}\} \rightarrow \{\beta, \alpha, -\vec{Q}\}$. A system with a center of inversion has symmetry with respect to $\{ \vec{Q}\} \rightarrow \{ -\vec{Q}\}$, in which case the tildes can be dropped for the diagonal components in $\{\alpha, \beta \}$ indices. This is the case for which the FDT is most frequently written \cite{Tranq:zali05}. The FDT, Eq.~(\ref{Tranq:Eq:FDT}), is a consequence of the detailed balance condition (\ref{Tranq:Eq:DelBal}) and the causality relations, which require that $\chi^{''}_{\alpha\beta} (\vec{Q}, \omega)$ is properly asymmetric. The fundamental laws of nature expressed in Eqs. (\ref{Tranq:Eq:DelBal}) and (\ref{Tranq:Eq:FDT}) are extremely useful in performing and analyzing neutron scattering experiments.

\subsection{Elastic and inelastic scattering}

If there exists a non-zero equilibrium magnetization in the sample, $\langle \vec{M}_{\vec{Q}} \rangle = \langle \overline{\vec{M}_{\vec{Q}} (t)} \rangle$, where the bar over $\vec{M}_{\vec{Q}} (t)$ denotes the time-averaging, we can introduce magnetization fluctuation around this equilibrium, $\vec{m}_{\vec{Q}} (t) = \vec{M}_{\vec{Q}} (t) - \langle \vec{M}_{\vec{Q}} \rangle$, and write
\begin{equation}%
S^{\alpha\beta} (\vec{Q}, \omega) = \langle M^{\alpha}_{\vec{Q}} \rangle \langle M^{\beta}_{-\vec{Q}} \rangle \delta(\hbar\omega) + S^{\alpha\beta}_{\rm inel } (\vec{Q}, \omega) \ts,
\label{Tranq:Eq:SMM-el-inel}\index{dynamic structure factor}
\end{equation}%
where the inelastic component $S^{\alpha\beta}_{\rm inel } (\vec{Q}, \omega)$ is defined similarly to Eq.~(\ref{Tranq:Eq:SMM-corr}), but with $\vec{M}_{\vec{Q}}$ replaced by $\vec{m}_{\vec{Q}}$. The first term here leads to elastic scattering which results from static magnetization in the sample, while the second term describes the inelastic magnetic scattering arising from its motion. Substituting the first term into Eq.~(\ref{Tranq:Eq:CrossSec-mag-S}) we obtain the unpolarized magnetic elastic cross-section,
\begin{equation}%
\frac{d^2 \sigma_{m,el} }{dE d\Omega} =  \frac{r_m^2}{(2 \mu_B)^2} \left| \langle M^{\perp}_{\vec{Q}} \rangle \right|^2 \delta(\hbar\omega) \ts,
\label{Tranq:Eq:CrossSec-mag-el}\index{Bragg diffraction, magnetic}
\end{equation}%
where $M^{\perp}_{\vec{Q}}$ is the Fourier transform of the magnetization component perpendicular to the wave vector transfer, $\vec{Q}$.

\subsection{Magnetic order and magnetic Bragg peaks}

Eq.~(\ref{Tranq:Eq:CrossSec-mag-el}) applies equally well to all cases where static magnetism is present in a crystal, whether it is a long-range magnetic order leading to Bragg peaks, or a short-range, e. g. nano-scale magnetic correlation, resulting in an appearance of a broad magnetic diffuse scattering. In the case of a long-range order, the magnetization density in a crystal typically has an equilibrium static component, which is modulated with a wave vector $\vec{Q}_m$,
\begin{equation}%
\langle \vec{M} (\vec{r}) \rangle = \vec{m}_0 (\vec{r}) + \vec{m} (\vec{r})\, {\rm e}^{i \vec{Q}_m\cdot\vec{r}} + \vec{m}^* (\vec{r})\, {\rm e}^{-i \vec{Q}_m\cdot\vec{r}} \ts,
\label{Tranq:Eq:M_r}
\end{equation}%
where $\vec{m}_0 (\vec{r})$ is a real vector function that describes the ferromagnetic component, if present, while  $\vec{m} (\vec{r})$ can be complex and describes the staggered magnetization. These ``Bloch amplitudes'' are periodic in the crystal lattice, and therefore can be expanded in the Fourier series,
\begin{equation}%
\vec{m} (\vec{r}) = \frac{1}{V_0} \sum_{\vec{\tau}} \vec{m}_{\vec{\tau}}\, {\rm e}^{i \vec{\tau}\cdot\vec{r}} , \;\;\;
\vec{m}_{\vec{\tau}} = \int_{V_0} \vec{m} (\vec{r})\, {\rm e}^{-i \vec{\tau}\cdot\vec{r}} \ts,
\label{Tranq:Eq:m_tau}
\end{equation}%
where the integral is over the unit cell of the nuclear (paramagnetic) crystal lattice.

Substituting (\ref{Tranq:Eq:M_r}) and (\ref{Tranq:Eq:m_tau}) into Eq.~(\ref{Tranq:Eq:CrossSec-mag-el}), we obtain the following expression for magnetic Bragg scattering associated with the long-range magnetic order at a wave vector $\vec{Q}_m$,
\begin{eqnarray}%
\frac{d^2 \sigma_{m,B} }{dE d\Omega} & = &  N r_m^2 V^* \sum_{\vec{\tau}} \left( \left| \frac{\vec{m}^{\perp}_{0,\vec{\tau}}}{2 \mu_B} \right|^2 \delta(\vec{Q} - \vec{\tau}) + \null \right. \nonumber\\
 & & \left. \left| \frac{\vec{m}^{\perp}_{\vec{\tau}}}{2 \mu_B} \right|^2 \left[ \delta(\vec{Q} - \vec{Q}_m + \vec{\tau}) + \delta(\vec{Q} + \vec{Q}_m + \vec{\tau}) \right] \right) \delta(\hbar\omega) \ts.
\label{Tranq:Eq:CrossSec-mag-el2}\index{Bragg diffraction, magnetic}
\end{eqnarray}%
Here the summation is over the paramagnetic crystal lattice. This is the ``large Brillouin zone'' description, which is the most general one, in that it does not rely on the existence of a commensurate magnetic superlattice with a unit cell containing some integer number of nuclear lattice unit cells, and applies to incommensurate, as well as commensurate magnetic structures. Such a description is most convenient for stripe phases in the cuprates, which are often incommensurate.

The intensities of magnetic satellites, $|\vec{m}^{\perp}_{\vec{\tau}}|^2$, are given by the Fourier amplitudes of the magnetization, (\ref{Tranq:Eq:m_tau}), which are obtained by performing the Fourier integrals over the unit cell of the paramagnetic lattice. In the case where the unit cell magnetization could be approximated by a number of point-like magnetic dipoles $\vec{\mu}_{\nu}$ located at positions $\vec{r}_{\nu}$, these amplitudes become the conventional unit cell magnetic structure factors,
\begin{equation}%
\vec{m} (\vec{r}) = \sum_{j,\nu} \vec{\mu}_{\nu}\, \delta \left( \vec{r} - \vec{R}_j - \vec{r}_\nu \right), \;\;\;
\left| \vec{m}^{\perp}_{\vec{\tau}} \right|^2 = \left| \sum_{\nu} \vec{\mu}^{\perp}_{\nu} {\rm e}^{ -i \vec{\tau}\cdot\vec{r}_\nu} \right|^2 \ts.
\label{Tranq:Eq:m_strfac}
\end{equation}%
In discussing magnetic scattering we assume a rigid lattice, neglecting atomic displacements due to disorder and vibrations discussed above. The leading correction to this description is obtained by multiplying expressions for magnetic cross-section with the Debye-Waller factor, ${\rm e}^{- 2W}$.

\subsection{Magnetic form factor and spin correlations}
\label{Tranq:sc:mff}\index{magnetic form factor}

In many important cases the magnetization density in the crystal is carried by electrons localized on atomic-like orbitals, which are specified by the local atomic variables, such as spin and orbital quantum numbers. In such cases, the matrix element of the atomic magnetization in the magnetic neutron scattering cross-section can be factorized into the product of the reduced matrix element (form factor), which does not depend on the direction of the atom's angular momentum quantum numbers, and the Wigner 3j-symbol, which entirely accounts for such dependence. Hence, the cross-section can be expressed in terms of a product of the $\vec{Q}-$dependent form factor, which accounts for the shape of the magnetization cloud associated with the atomic spin and orbital variables, and a dynamical correlation function between these local angular momentum variables at different lattice sites.

For magnetic ions obeying Hund's rule, neutron scattering usually probes states belonging to the same multiplets of the angular momentum, $\Delta L = 0$, $\Delta S = 0$ for the Russel-Saunders atoms with weak spin-orbit and strong crystal field, or $\Delta J = 0$ for the case of strong spin-orbit coupling, where the total angular momentum $\vec{J} = \vec{L} + \vec{S}$ is a good quantum number, such as in rare earths. Hence, we can write for the Fourier transform of atomic magnetization,
\begin{equation}%
\langle \eta_f | \vec{M} (\vec{Q}) |\eta_i \rangle = -2\mu_B F_S (\vec{Q}) \langle \eta_f | \vec{S} |\eta_i \rangle - \mu_B F_L (\vec{Q}) \langle \eta_f | \vec{L} |\eta_i \rangle \ts,
\label{Tranq:Eq:SO_WE}
\end{equation}%
where the spin and the orbital magnetic form factors are,
\begin{equation}%
F_S (\vec{Q}) = \frac{\langle \eta'_f, L, S | \sum_e {\rm e}^{-i \vec{Q}\cdot\vec{r}_e} \left( \vec{s}_e \cdot \vec{S} \right) | \eta'_i, L, S \rangle }{S(S+1)} \ts,
\label{Tranq:Eq:S_FF}\index{magnetic form factor}
\end{equation}%
\begin{equation}%
F_L (\vec{Q}) = \frac{\langle \eta'_f, L, S | \sum_e {\rm e}^{-i \vec{Q}\cdot\vec{r}_e} \left( \vec{\mu}_{e,l} \cdot \vec{L} \right) | \eta'_i, L, S \rangle }{\mu_B L(L+1)} \ts,
\label{Tranq:Eq:O_FF}
\end{equation}%
where we made explicit that initial and final states of the sample belong to the same $L$ and $S$ multiplet. Similar relations hold for the $J$ multiplet in the strong spin-orbit coupling limit. 

Typically it is possible to define an effective spin operator,
\begin{equation}%
\langle \eta_f | \vec{M} (\vec{Q}) |\eta_i \rangle = - g \mu_B F (\vec{Q}) \langle \eta_f | \tilde{\vec{S}} |\eta_i \rangle \ts,
\label{Tranq:Eq:S_WE}
\end{equation}%
\begin{equation}%
F (\vec{Q}) = \frac{\langle \eta'_f, L, S |  \left( \vec{M} (\vec{Q}) \cdot \tilde{\vec{S}} \right) | \eta'_i, L, S \rangle }{g\mu_B \tilde{S} (\tilde{S}+1)} = \frac{g_S}{g} F_S (\vec{Q}) + \frac{g-g_S}{g} F_L (\vec{Q})
\ts.
\label{Tranq:Eq:Seff_FF}
\end{equation}%
where $g$ and $g_S$ are the effective $g-$factors, $\langle \eta_f | \vec{L} + 2 \vec{S} |\eta_i \rangle = g \langle \eta_f | \tilde{\vec{S}} |\eta_i \rangle$, $\langle \eta_f | 2 \vec{S} |\eta_i \rangle = g_S \langle \eta_f | \tilde{\vec{S}} |\eta_i \rangle$. These expressions are exact in the cases of a $J$-multiplet, where $\tilde{\vec{S}} = \vec{J}$, $g_S = 2(g - 1)$ and $g_L = 2 - g$, or a pure spin multiplet, where $g = g_S$ and the orbital contribution is absent. They give the leading-order approximation in other cases. If the orbital moment is nearly quenched, as it is for magnetic $d$-elements in strong crystal field, then $\tilde{\vec{S}} \approx \vec{S}$, $g_S \approx 2$, and the orbital contribution to $F(\vec{Q})$, is small. Assuming this to be the case, we shall omit tildes and use $\vec{S}$ for the effective spin.

Using the factorization of atomic magnetization provided by Eqs. (\ref{Tranq:Eq:S_WE}) and (\ref{Tranq:Eq:Seff_FF}), the magnetic neutron scattering cross-section (\ref{Tranq:Eq:CrossSec-mag}) can be recast as
\begin{eqnarray}%
\frac{d^2 \sigma_{m} }{dE d\Omega} = \frac{k_f}{k_i} r_m^2 \sum_{\alpha,\beta} \left(\delta_{\alpha \beta} - \frac{Q_{\alpha} Q_{\beta}}{Q^2} \right) \sum_{j,j'} g_{\alpha,j} \frac{F^*_j (\vec{Q})}{2} g_{\beta,j'} \frac{F_{j'} (\vec{Q})}{2} \times \nonumber\\
\int_{-\infty}^{\infty} {\rm e}^{ -i\omega t } {\rm e}^{ - i \vec{Q}\cdot(\vec{R}_{j} - \vec{R}_{j'}) } \langle S^\alpha_{j} S^\beta_{j'} (t) \rangle \frac{dt}{2 \pi \hbar} \ts, \;\;\;
\label{Tranq:Eq:CrossSec-spin}
\end{eqnarray}%
where we allow for the possibility that the $g-$factor is anisotropic, and that both $g_{\alpha,j}$ and $F_{j} (\vec{Q})$ could be different for different sites $j, j'$ of the lattice. Equation (\ref{Tranq:Eq:CrossSec-spin}) relates the magnetic cross-section to the dynamic spin structure factor, which is the Fourier transform of the time-dependent two-point correlation function of the atomic spin variables on the sites of the lattice,
\begin{equation}%
S^{\alpha\beta} (\vec{Q}, \omega) = \int_{-\infty}^{\infty} {\rm e}^{ - i\omega t } \frac{1}{N} \sum_{j,j'} {\rm e}^{ - i \vec{Q} (\vec{R}_{j} - \vec{R}_{j'}) } \langle S^\alpha_{j} S^\beta_{j'} (t) \rangle \frac{dt}{2 \pi \hbar} \ts.
\label{Tranq:Eq:S-SS-corr}\index{dynamic structure factor}
\end{equation}%
$S^{\alpha\beta} (\vec{Q}, \omega)$ is a quantity which is calculated in theoretical models based on the local spin Hamiltonians. It also obeys a number of important relations, known as sum rules, which are extremely useful in analyzing neutron scattering data. The zero moment sum rule is obtained by integrating Eq.~(\ref{Tranq:Eq:S-SS-corr}) in $\vec{Q}$ and $\omega$, providing the direct connection of the integral neutron intensity with the spin value $S$ in the lattice spin Hamiltonian,
\begin{equation}%
\sum_{\alpha} \int_{-\infty}^{\infty} S^{\alpha\alpha} (\vec{Q},\omega) d^3 \vec{q} d(\hbar\omega) = S (S+1) \ts.
\label{Tranq:Eq:S-SS-sum}\index{dynamic structure factor, sum rule}
\end{equation}%
The first moment sum rule relates $\sum_{\alpha} \int_{-\infty}^{\infty} \hbar\omega S^{\alpha\alpha} (\vec{Q},\omega) d^3 \vec{q} d(\hbar\omega)$, which is the integral oscillator strength of the fluctuation spectrum, with the bond energies in the spin Hamiltonian, and so on.

\subsection{Spin waves}
\index{spin waves}

Representing the neutron scattering cross-section via two-point dynamical spin correlation function, as in Eq. (\ref{Tranq:Eq:CrossSec-spin}), is possible in a large number of important magnetic systems, such as cuprates and other $3d$ magnetic insulators. Such a representation is extremely useful, as it allows one to connect the measured magnetic neutron intensity with the theoretically predicted properties of model spin Hamiltonians, such as the Heisenberg spin Hamiltonian,

\begin{equation}%
\hat{\cal{H}} = \sum_{j,j'} J_{jj'} \vec{S}_{j} \vec{S}_{j'} = \sum_{\vec{q}} N J_{\vec{q}} \vec{S}_{\vec{q}} \vec{S}_{-\vec{q}} \ts.
\label{Tranq:Eq:Heis_Spin_Hamilt}
\end{equation}%
Here $J_{jj'} = J(\vec{r}_{jj'})$ is the exchange coupling between sites $j$ and $j'$, and $J_{\vec{q}}$ and $\vec{S}_{\vec{q}}$ are the lattice Fourier transforms,

\begin{equation}%
J_{\vec{q}} = \sum_{\vec{r}_{jj'}} J_{jj'} {\rm e}^{-i \vec{q}\cdot\vec{r}_{jj'}} , \;\;\; \vec{S}_{\vec{q}} = \sum_{j} \vec{S}_{j} {\rm e}^{-i \vec{q}\cdot\vec{r}_{j}} \ts.
\label{Tranq:Eq:SqJq}
\end{equation}%

In many systems with magnetic order, the average value of spin at each lattice site in the ground state (GS) is ``frozen'' at nearly the full saturation value, $\langle S_j^z \rangle \approx S$. In particular, this is a very good approximation for the semi-classical spins, $S \gg 1$, in more than one dimension (1D). For quantum spins, $S = 1/2$, and/or in the low-dimensional, or frustrated systems, the order may be weak, or absent, and such a picture is inadequate. Nevertheless, in a large number of systems magnetic order in the ground state is well developed, and the semiclassical spin-wave picture applies.

Spin excitations in a magnetic system with a well-ordered ground state, such as a ferromagnet, where all spins are parallel, or a semi-classical antiferromagnet, where there are two antiparallel sublattices, can be visualized as small oscillations of classical spin vectors around their equilibrium positions in the GS spin structure. Their wave-like spatial composition results from the translational symmetry of the system. Frequencies of such spin-wave oscillations can be calculated from the spin Hamiltonian, such as (\ref{Tranq:Eq:Heis_Spin_Hamilt}), totally within the classical mechanics, simply by writing the torque equations of motion for the classical spin angular momenta. For example, in the case of the Heisenberg Hamiltonian (\ref{Tranq:Eq:Heis_Spin_Hamilt}) for a magnetically ordered system characterized by the ordering wave vector $\vec{Q}_0$ (this includes ferromagnetism with $\vec{Q}_0 = 0$, as well as antiferromagnetism and helimagnetism), one obtains the spin-wave dispersion \cite{Tranq:vill59,Tranq:yosh59},
\begin{equation}%
\hbar\omega_{\vec{q}} = 2S \sqrt{ \left( J_{\vec{q}} - J_{\vec{Q}_0} \right) \left( \frac{J_{\vec{q}+\vec{Q}_0} + J_{\vec{q}-\vec{Q}_0}}{2} - J_{\vec{Q}_0} \right) } \ts.
\label{Tranq:Eq:SW_disp}\index{spin waves}\index{spin-wave dispersion}
\end{equation}%
This can be recast as $\omega_{\vec{q}} = \sqrt{ \omega_{\vec{0}} \omega_{\vec{Q}_0} }$, where $\hbar\omega_{\vec{0}} = 2S (J_{\vec{q}} - J_{\vec{Q}_0} )$. 

Spin waves are the normal modes of the linearized equations of motion. They involve small spin deviations that are perpendicular to the equilibrium spin direction. Hence, spin waves are transversely polarized, with two mutually orthogonal linear polarizations of spin oscillations possible. For a spin system on a Bravais lattice there are two spin-wave modes. 

In a quantum-mechanical treatment of spins, the spin-wave calculation proceeds via an approximate mapping of spin operators to Bose creation-annihilation operators, {\it i.e.} to local oscillator modes. Hence, the so obtained spin-wave theory (SWT) describes spin excitations as coherent waves of small oscillations around the local equilibrium positions, in many ways similar to phonons. The resulting expression for the spin-wave contribution to the neutron magnetic scattering cross-section in a sample with a spiral spin structure with the propagation vector $\vec{Q}_0$ is
\begin{eqnarray}
\frac{d^2 \sigma_{sw}}{dE d\Omega}  = \frac{k_f}{k_i} r_m^2 N \left| \frac{g}{2} F(\vec{Q}) \right|^2 \frac{S}{2} V^* \sum_{\vec{\tau}} (n(\omega) + 1) \delta(\hbar\omega - \hbar\omega_{\vec{q}}) \times \;\;\; \nonumber\\
\left[ \frac{1}{4} \left(1 + \frac{Q_z^2}{Q^2} \right) \sqrt{ \frac{\omega_{\vec{0}} }{ \omega_{\vec{Q}_0} }} \left( \frac{}{} \delta(\vec{Q} - \vec{q}- \vec{\tau} - \vec{Q}_0) + \delta(\vec{Q} - \vec{q}- \vec{\tau} + \vec{Q}_0) \right) \right. +  \nonumber\\
\left. \left(1 - \frac{Q_z^2}{Q^2} \right) \sqrt{ \frac{\omega_{\vec{Q}_0} }{ \omega_{\vec{0}} }} \delta(\vec{Q} - \vec{q} - \vec{\tau}) \right] , \;\;\;
\label{Tranq:Eq:CrossSec-SW-spiral}\index{spin waves, neutron scattering cross section}
\end{eqnarray}%
where $z$ is the direction normal to the plane of the spiral, and we have restricted consideration to the case of a Bravais lattice and retained only the contribution corresponding to creation of a single spin wave. The contribution arising from the absorption of a spin wave is written similarly to that of a phonon in Eq. (\ref{Tranq:Eq:CrossSec-phonon-Brav}). For a ferromagnet, the single spin-wave magnetic cross-section simplifies to, \cite{Tranq:squi12,Tranq:love84},
\begin{eqnarray}
\frac{d^2 \sigma_{sw} }{dE d\Omega} & =&  \frac{k_f}{k_i} r_m^2 N \left| \frac{g}{2} F(\vec{Q}) \right|^2 \frac{S}{2} \left(1 + \frac{\vec{Q}_\parallel^2 }{Q^2} \right) \times \;\;\; \nonumber\\
 & & \quad V^* \sum_{\vec{\tau}} \left[ \frac{}{} \delta(\vec{Q} - \vec{q} - \vec{\tau}) \delta(\hbar\omega - \hbar\omega_{\vec{q}}) (n(\omega) + 1)  \right. +  \nonumber\\
 & & \qquad\qquad\quad \left. \delta(\vec{Q} + \vec{q} - \vec{\tau}) \delta(\hbar\omega + \hbar\omega_{\vec{q}}) n(\omega) \frac{}{} \right] , \;\;\;
\label{Tranq:Eq:CrossSec-SW-FM}\index{spin waves, neutron scattering cross section}
\end{eqnarray}%
where $\vec{Q}_{\parallel}$ is the wave vector component along the ferromagnetic ordered moment and we have retained the contributions from both the creation and the absorption of a spin wave.

\subsection{Anisotropic magnetic form factor and covalency}
\index{magnetic form factor, anisotropy}\index{magnetic form factor, covalency}

It is clear from Eq.~(\ref{Tranq:Eq:CrossSec-spin}) that even the exact knowledge of the dynamical spin structure factor (available from theory in some special cases, such as in one dimension) is insufficient to reproduce the measured magnetic scattering cross-section. One also has to know the magnetic form factor, which needs to be obtained from an ab initio calculation of the electronic density in the crystal. 

In the most common case of a Hund's ion with $2S$ unpaired electrons forming spin $(2S+1)$-multiplet, the spin magnetic form factor (\ref{Tranq:Eq:S_FF}) becomes
\begin{equation}%
F_S (\vec{Q}) = \frac{1}{2S} \sum_{e=1}^{2S} \int {\rm e}^{-i \vec{Q}\cdot\vec{r}} \left| \psi_e (\vec{r}) \right|^2 d^3 \vec{r} = 
\frac{1}{2S} \sum_{e=1}^{2S} F_{S,e} (\vec{Q}) \ts,
\label{Tranq:Eq:S_FF_Hund}
\end{equation}%
where the sum is only over the unpaired electrons. The single-electron density, $|\psi_e (\vec{r})|^2$, is determined from the many-electron atomic wave function through $|\psi_e (\vec{r})|^2 = \langle \eta', L, S | \delta (\vec{r} - \vec{r}_e) | \eta', L, S \rangle$. The magnetic form factor for an atom is therefore simply an average of those for each of the unpaired electrons. Similarly, the orbital form factor is the Fourier-transformed average density of the uncompensated orbital currents in the atom. 

If the average Hartree-Fock potential acting on an unpaired electron $e$ in the atom is spherically symmetric, then the effective one-electron wave functions in (\ref{Tranq:Eq:S_FF_Hund}) are the eigenfunctions of angular momentum and are tagged by the $n, l, m = l^z$ quantum numbers, $\psi_e (\vec{r}) = \psi_{n,l,m} (\vec{r})$. The angular and the radial dependencies of the electronic density factorize, $\psi_{n,l,m} (\vec{r}) = R_{n,l} (r) Y_l^m (\theta, \phi)$, where $Y_l^m (\theta, \phi)$ is the spherical function giving the dependence on the polar angles $\theta, \phi$. This so-called central field approximation is good when the contribution to the potential from electrons in the incomplete shell is small. However, it also becomes exact for an almost-filled shell, with only a single electron, or a single hole, as in the case of Cu$^{2+}$, or for a nearly half-filled shell, because the average potential of the closed, or half-filled shell, is spherically symmetric. 

In the general case, a single-electron wave function can always be expanded in a series in spherical harmonics. In each term of such an expansion, the radial and the angular parts are again factorized, and the magnetic form factor is a sum of Fourier-transformed terms with different $l$ and $m$. The same kind of an expansion is encountered in calculating the orbital contribution to the magnetic form factor. This is known as a multipole expansion \cite{Tranq:love84}. The calculations are ion-specific and extremely cumbersome. The general expressions can be obtained only for the leading, isotropic contributions, in the limit of small wave vector transfer, known as the dipole approximation,
\begin{equation}%
F_S (\vec{Q}) = \langle j_0 (Q) \rangle \, , \;\;\; F_L (\vec{Q}) = \frac{1}{2} \left( \langle j_0 (Q) \rangle + \langle j_2 (Q) \rangle \right) \ts,
\label{Tranq:Eq:S_FF_dipole}
\end{equation}%
where $j_0 (Q)$ and $j_2 (Q)$ are the $l, m$ dependent radial integrals quantifying the radial wave function \cite{Tranq:inte95,Tranq:zali05}. The radial integrals for most known magnetic atoms and ions have been calculated numerically from the appropriate Hartree-Fock or Fock-Dirac wave functions and are tabulated in Ref \cite{Tranq:inte95}. The full $F (\vec{Q})$ is given by Eq.~(\ref{Tranq:Eq:Seff_FF}).

\index{magnetic form factor, Ir$^{4+}$}\index{K$_2$IrCl$_6$}

Although the dipole approximation (\ref{Tranq:Eq:S_FF_dipole}) is the one most commonly used, it is extremely crude. In particular, it does not account for the anisotropy of the magnetic form factors, which can be very important for ions with only one or two unpaired electrons. The anisotropic magnetic form factor of a single $5d$ hole in a $t_{2g}$ orbital of the magnetic Ir$^{4+}$ ion in the cubic K$_2$IrCl$_6$ was studied in Ref.~\cite{Tranq:lynn76}. The authors found that the anisotropy of the magnetic form factor is very large, with an additional enhancement coming from the hybridization of the Ir $5d$-orbital with the Cl $p$-orbitals. 

\index{magnetic form factor, Cu$^{2+}$}\index{La$_2$CuO$_4$}\index{YBa$_2$Cu$_3$O$_{6+y}$}
\index{SrCuO$_2$}\index{Sr$_2$CuO$_3$}

The anisotropy of the magnetic form factor is also very pronounced in La$_2$CuO$_4$, YBa$_2$Cu$_3$O$_{6+y}$, and related cuprate materials, including the high-$T_c$ superconductors, where in the ionic picture a single unpaired magnetic electron occupies a $3d_{x^2-y^2}$ orbital. In Ref. \cite{Tranq:sham93} the authors found that properly accounting for the anisotropy of the Cu$^{2+}$ magnetic form factor is essential for understanding the magnetic Bragg intensities measured in YBa$_2$Cu$_3$O$_{6+y}$ at large wave vectors, and can also explain the peculiar $\vec{Q}$-dependence of the inelastic magnetic cross-section in this
material. Accounting for the anisotropic Cu$^{2+}$ form factor was also very important in analyzing neutron scattering by high-energy spin waves in
La$_2$CuO$_4$ \cite{Tranq:cold01,Tranq:head10}, and the chain cuprates SrCuO$_2$ and Sr$_2$CuO$_3$ \cite{Tranq:zali04,Tranq:walt09}. The magnetic excitations in these cuprate materials extend to several hundreds of meV. Consequently, the measurements require very large wave vector transfers, for which the anisotropy of the magnetic form factor is very pronounced.

The ionic magnetic form factors for $3d$ orbitals can be explicitly computed by Fourier transforming the corresponding spherical harmonics. In particular, for the $d_{x^2-y^2}$ orbital relevant for Cu$^{2+}$ one obtains \cite{Tranq:zali05},
\begin{eqnarray}%
F (\vec{Q}) & = & \langle j_0 (Q) \rangle - \frac{5}{7} \langle j_2 (Q) \rangle \left( 1- \cos^2 \theta_{\vec{Q}} \right) \null \nonumber \\
 & & \null + \frac{9}{56} \langle j_4 (Q) \rangle \left( 1 - 10 \cos^2 \theta_{\vec{Q}} + \frac{35}{3} \cos^4 \theta_{\vec{Q}} \right) \nonumber\\
 & & \null + \frac{15}{8} \langle j_4 (Q) \rangle \sin^4 \theta_{\vec{Q}} \cos \left(4 \phi_{\vec{Q}} \right) \ts, 
\label{Tranq:Eq:Cu-3d-FF}
\end{eqnarray}%
where $\theta_{\vec{Q}}, \phi_{\vec{Q}}$ are the polar angles of the wave vector $\vec{Q}$ in the local coordinate system used to specify the proper orbital wave functions in the crystal field.

\begin{figure}[t]
\centering
\includegraphics*[width=1.\textwidth]{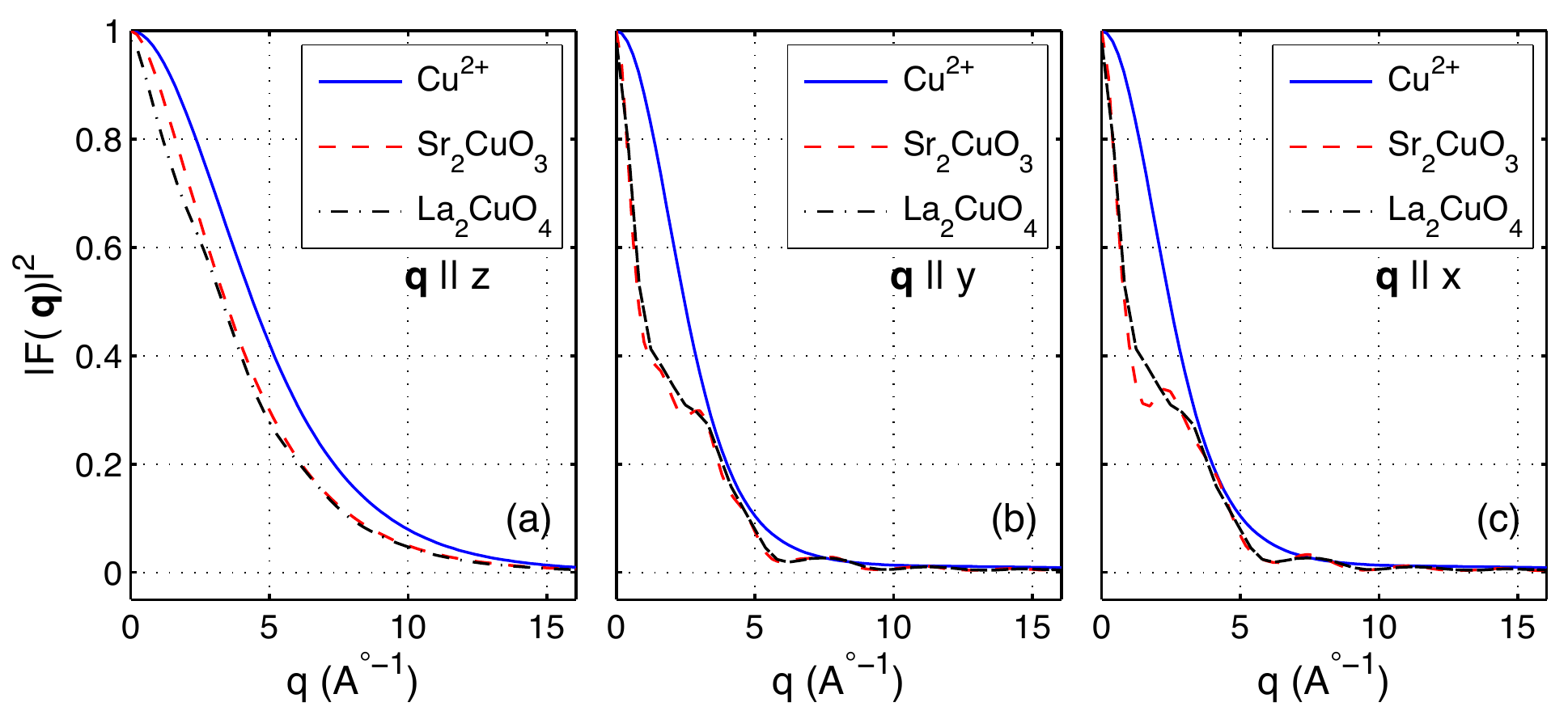}
\caption[]{Wave vector dependence of the ionic magnetic form factor of Cu$^{2+}$ given by Eq.~(\ref{Tranq:Eq:Cu-3d-FF}) (solid line) and the covalent magnetic form factors for Sr$_2$CuO$_3$ (dashed line) and La$_2$CuO$_4$ (dash-dotted line) obtained from the ab initio density functional calculations \cite{Tranq:walt09}. Panels (a) - (c) show the dependence along three principal directions.}
\label{Tranq:Fig:FF2_Cu}\index{magnetic form factor, Cu$^{2+}$}\index{Sr$_2$CuO$_3$} \index{La$_2$CuO$_4$}
\end{figure}

Although using the anisotropic ionic magnetic form factor of Cu$^{2+}$ is much better than using a spherical form factor of the dipole approximation, it is still not sufficient for cuprates, as it neglects the effects of covalency (i.e., charge transfer to the neighboring oxygen) that are expected to be very significant in these materials. In Ref. \cite{Tranq:walt09} it was discovered that covalent bonding results in a marked modification of the magnetic form factor in the quasi-1D antiferromagnet Sr$_2$CuO$_3$. The local structure of the planar Cu--O square plaquettes in this material is essentially identical to that in La$_2$CuO$_4$. Making use of a precise theoretical result for the excitation spectrum available in 1D, the authors demonstrated that a good fit to the data requires a form factor that takes account of hybridization between the half-filled Cu $3d_{x^2-y^2}$ orbital and the ligand O $2p_\sigma$ orbitals, as given by a density functional calculation. The hybridization causes the spin density to be extended in real space, resulting in a more rapid fall off in reciprocal space compared to a simple Cu$^{2+}$ form factor, as illustrated in Fig.~\ref{Tranq:Fig:FF2_Cu}. Smaller values of magnetic form factor at relatively large wave vectors, where the measurement is performed, lead to the suppression of magnetic intensity, which could be as large as a factor of two or more \cite{Tranq:walt09}. Finally, we note that a study of covalent NMR shifts by Walstedt and Cheong \cite{Tranq:wals01} found that barely 2/3 of the spin density in La$_2$CuO$_4$ resides on the copper sites, in excellent agreement with the Sr$_2$CuO$_3$ neutron data of Walters {\it et al.} \cite{Tranq:walt09}.

\index{La$_2$CuO$_4$}\index{Sr$_2$CuO$_3$}

\section{Application to cuprate superconductors}
\index{cuprate superconductors}\index{La$_{2-x}$Ba$_x$CuO$_4$}\index{antiferromagnetic order}

\begin{figure}[t]
\centering
\includegraphics*[width=.7\textwidth]{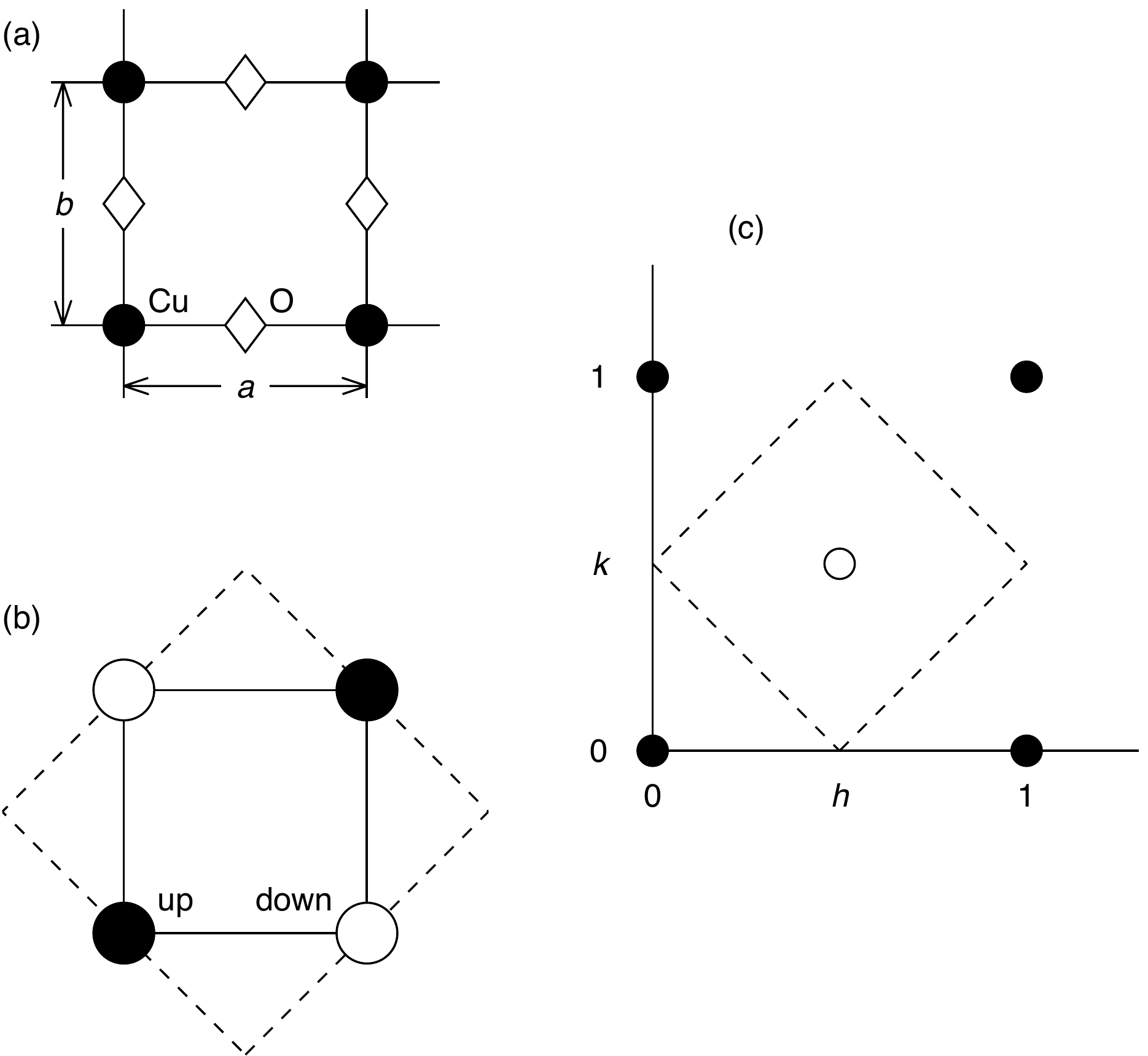}
\caption[]{(a) Structure of a CuO$_2$ plane, with Cu atoms indicated by filled circles and O atoms by open diamonds.  (b) Schematic of antiferromagnetic order, with alternating up (filled circles) and down (open circles) spins.  The solid line indicates the chemical unit cell, while the dashed line indicates the doubled area of the antiferromagnetic unit cell.  (c) Reciprocal space showing fundamental Bragg peak positions (filled circles) and antiferromagnetic superlattice peak (open circle) at $(\frac12,\frac12)$.}
\label{Tranq:fg:AF}
\end{figure}

The discovery of high-temperature superconductivity in La$_{2-x}$Ba$_x$CuO$_4$\linebreak 
(LBCO) came as a considerable surprise \cite{Tranq:bedn86}, as ceramic oxides were generally considered to be poor conductors.  The structure of LBCO and related cuprates involves CuO$_2$ layers, with the Cu atoms forming a square lattice with bridging O atoms, as shown in Fig.~\ref{Tranq:fg:AF}(a).  Anderson \cite{Tranq:ande87} predicted that the parent compound, La$_2$CuO$_4$, should have strong antiferromagnetic (AF) superexchange interactions between nearest-neighbor Cu atoms.  The occurrence of antiferromagnetic order was demonstrated by Vaknin {\it et al.} \cite{Tranq:vakn87} using neutron diffraction on a powder sample of La$_2$CuO$_4$.  As illustrated in Fig.~\ref{Tranq:fg:AF}(b), the antiferromagnetic N\'eel order doubles the size of the unit cell in real space, which results in magnetic superlattice peaks, as shown in (c).  Thus, the antiferromagnetic order can be detected through the appearance of superlattice peaks.  The challenge in this case is that one must distinguish from structural superlattice peaks due to staggered rotations of CuO$_6$ octahedra \cite{Tranq:axe94}.  Fortunately, the AF and structural peaks appear at inequivalent positions.

\begin{figure}[t]
\centering
\includegraphics*[width=.4\textwidth]{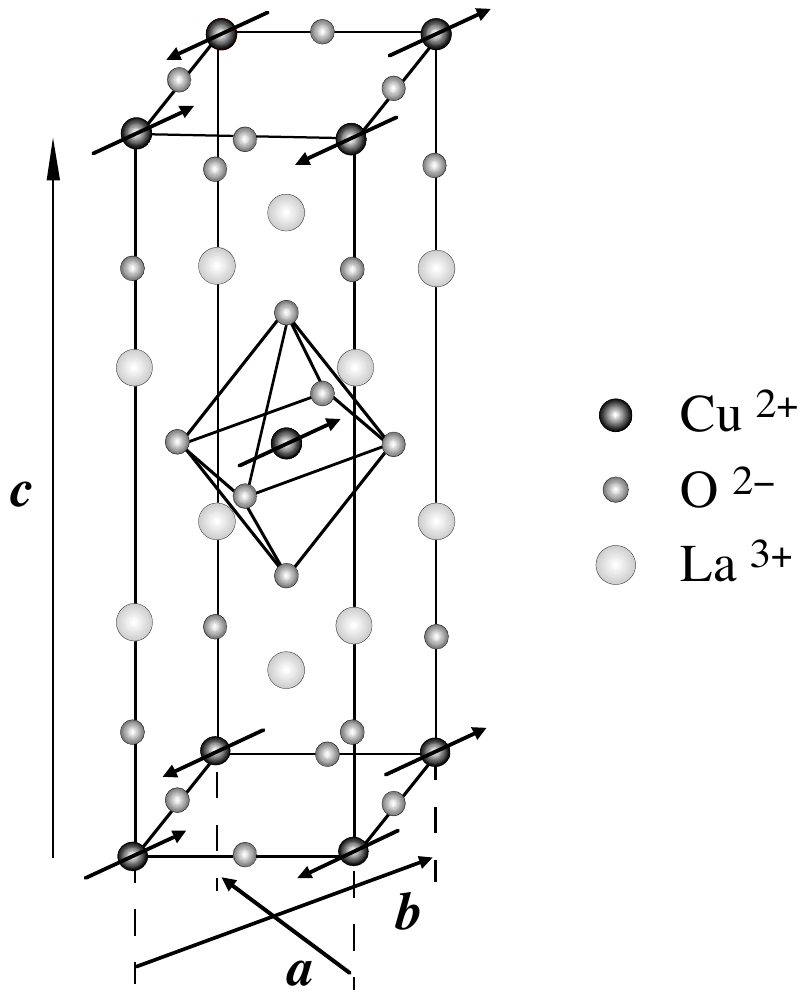}
\caption[]{Structure of La$_2$CuO$_4$, with arrows indicating the arrangement of the magnetic moments in the antiferromagnetic state.  Figure reprinted with permission from Lee {\it et al.} \cite{Tranq:lee99}. Copyright (1999) by the American Physical Society.}
\label{Tranq:fg:struc}\index{La$_2$CuO$_4$, antiferromagnetic order}
\end{figure}

The ordered pattern of the octahedral tilts is associated with an orthorhombic distortion of the crystal structure that makes the diagonal directions of a Cu-O plaquette inequivalent \cite{Tranq:axe94}, as indicated in Fig.~\ref{Tranq:fg:struc}.  By analyzing the $\vec{Q}$ dependence of the AF Bragg peak intensities, it was possible to determine that the magnetic moments on Cu atoms lie within the CuO$_2$ planes, pointing along the orthorhombic $b$ axis \cite{Tranq:vakn87}.  Furthermore, it was possible to show that the relative arrangement in neighboring planes is as shown in Fig.~\ref{Tranq:fg:struc}.  With the magnetic structure determined, one can evaluate the magnitude of the magnetic moments by normalizing the AF peak intensities to the nuclear intensities and correcting for the magnetic form factor.  Early studies yielded a small ordered moment whose magnitude was correlated with the magnetic ordering, or N\'eel, temperature, $T_{\rm N}$ \cite{Tranq:yama87}.   Neutron scattering studies on carefully prepared single-crystal samples eventually demonstrated the impact of interstitial oxygen, within the La$_2$O$_2$ layers \cite{Tranq:well97}.  Removing the excess oxygen by annealing, one can achieve $T_{\rm N}=325$~K \cite{Tranq:keim92a} and a magnetic moment of $0.60\pm0.05$~$\mu_{\rm B}$ \cite{Tranq:yama87}.

\index{La$_2$CuO$_4$, antiferromagnetic order}

If one assumes a $g$ factor of roughly 2, then the ordered moment yields an average ordered spin $\langle S\rangle\approx 0.3$, compared to the expected $S=\frac12$ per Cu atom.  The reduction results from the strongly anisotropic structure and the low value of the spin.  For a two-dimensional magnetic system described by a Heisenberg spin Hamiltonian, long-range order is destroyed at any finite temperature by thermal excitation of spin fluctuations.  For La$_2$CuO$_4$, weak (nearly-frustrated) couplings between the planes enable the ordering at finite temperature \cite{Tranq:kast98}.  Nevertheless, the spin correlations have a strongly two-dimensional (2D) character, as demonstrated by neutron scattering studies \cite{Tranq:shir87}.  The small magnitude of the spin, combined with the enhanced zero-point spin fluctuations in 2D, puts the system close to a quantum critical point \cite{Tranq:chak88}.  Although the large fluctuations cause problems for perturbation theory, spin-wave theory nonetheless yields a result, $\langle S\rangle=0.3$, that is very close to the value obtained from experiment \cite{Tranq:mano91}.

\index{superexchange energy}\index{La$_2$CuO$_4$, spin waves}

\begin{figure}[t]
\centering
\includegraphics*[width=0.6\textwidth]{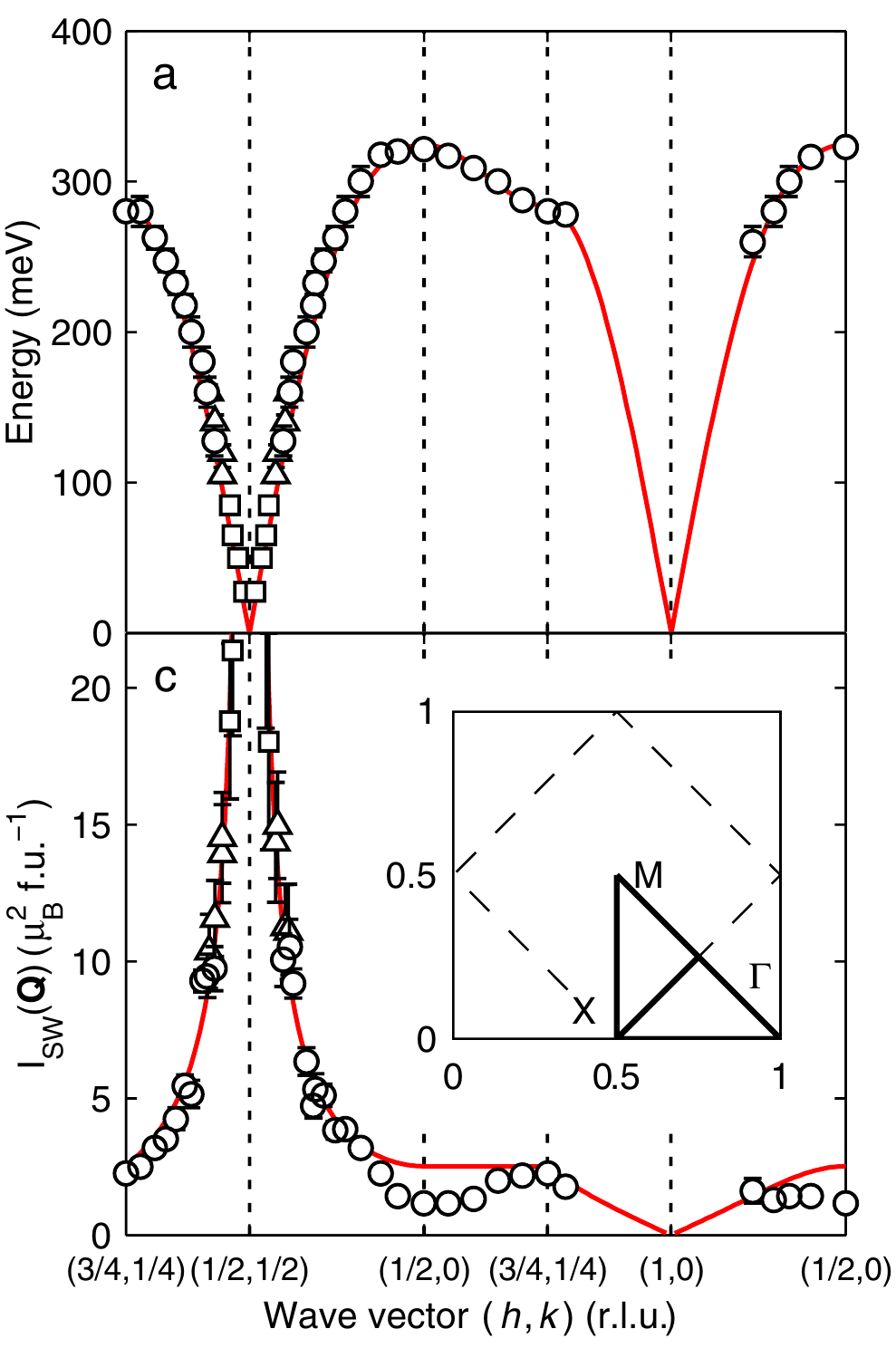}
\caption[]{Spin wave (a) dispersion and (c) intensity measured in antiferromagnetic La$_2$CuO$_4$ at $T=10$~K.  Lines through data correspond to fits with spin-wave theory; the fit to the intensity includes a renormalization factor $Z_d=0.4\pm0.04$.  Figure reprinted with permission from Headings {\it et al.} \cite{Tranq:head10}. Copyright (2010) by the American Physical Society.}
\label{Tranq:fg:lco}\index{La$_2$CuO$_4$, spin wave dispersion}
\end{figure}

The exchange couplings between the spins can be determined by analyzing the dispersion of the spin excitations, which can be obtained by inelastic scattering measurements on a single-crystal sample.  Early studies of La$_2$CuO$_4$ with triple-axis spectrometers demonstrated that the superexchange energy $J$ coupling nearest-neighbor spins is greater than 100~meV, and that effects such as exchange anisotropy and interlayer coupling are very small \cite{Tranq:kast98}.  Time-of-flight techniques were required to measure the highest-energy spin waves \cite{Tranq:hayd91}, and these have been refined over time \cite{Tranq:cold01,Tranq:head10}.  The most recent results, from Headings {\it et al.} \cite{Tranq:head10}, are shown in Fig.~\ref{Tranq:fg:lco}.  The line through the data points corresponds to a fit with linear spin-wave theory, which works surprisingly well in light of the the large zero-point fluctuations.  One impact of the latter is the renormalization factor, with a fitted value of $Z_d=0.4\pm0.04$, that is required to fit the measured intensity.  This is somewhat smaller than the value $Z_d\approx 0.6$ that is predicted from quantum corrections to linear spin waves \cite{Tranq:lore05}.  It should be noted, though, that this analysis did not take account of the hybridization effects on the magnetic form factor, discussed in Sec.~\ref{Tranq:sc:mff}, which would account for some of the apparent renormalization.  An additional effect is the damping and broadening of the energy-dependence of the spin-wave line shape at the zone boundary position $\vec{Q} = (\frac12,0)$.  This appears to be the result of interactions with a multi-magnon high-energy continuum \cite{Tranq:head10}.

The fitted dispersion corresponds to $J=143\pm2$~meV, but also requires longer-range exchange couplings---second and third neighbor couplings $J'$ and $J''$, which are relatively weak, and a significant 4-spin cyclic exchange term $J_c\approx 0.4J$.  The overall bandwidth of the magnetic spectrum is $\sim2J$.  A recent analysis of the couplings, including $J_c$, in terms of a single-band Hubbard model has been given by Dalla Piazza {\it et al.} \cite{Tranq:dall12}.

\index{La$_2$CuO$_4$, spin waves}

To achieve superconductivity, one must dope charge carriers into the CuO$_2$ planes.  Substituting Ba$^{2+}$ or Sr$^{2+}$ for La$^{3+}$ introduces holes.   A small density of holes, $p\approx2$\%, is enough to kill the long-range AF order, which is followed by a regime of spin-glass order \cite{Tranq:birg06}.  Doping beyond $p\sim0.055$ yields superconductivity.  The maximum superconducting transition temperature $T_c$ occurs for $p\sim0.16$, with $T_c$ heading towards zero for $p>0.25$.  Inelastic neutron scattering studies have been performed on single crystal samples across this entire doping range \cite{Tranq:birg06,Tranq:fuji12a}.  A couple of the key features are: 1) the bandwidth of strong spin-fluctuation scattering decreases linearly with doping, being quantitatively similar to the pseudogap energy extracted from various electron spectroscopies \cite{Tranq:stoc10,Tranq:fuji12a}, and 2) the wave vector characterizing the low-energy spin excitations splits about the AF wave vector, becoming incommensurate \cite{Tranq:yama98a,Tranq:birg06}.

\index{La$_{1.48}$Nd$_{0.4}$Sr$_{0.12}$CuO$_4$}\index{incommensurate magnetic order}

\begin{figure}[t]
\centering
\includegraphics*[width=.7\textwidth]{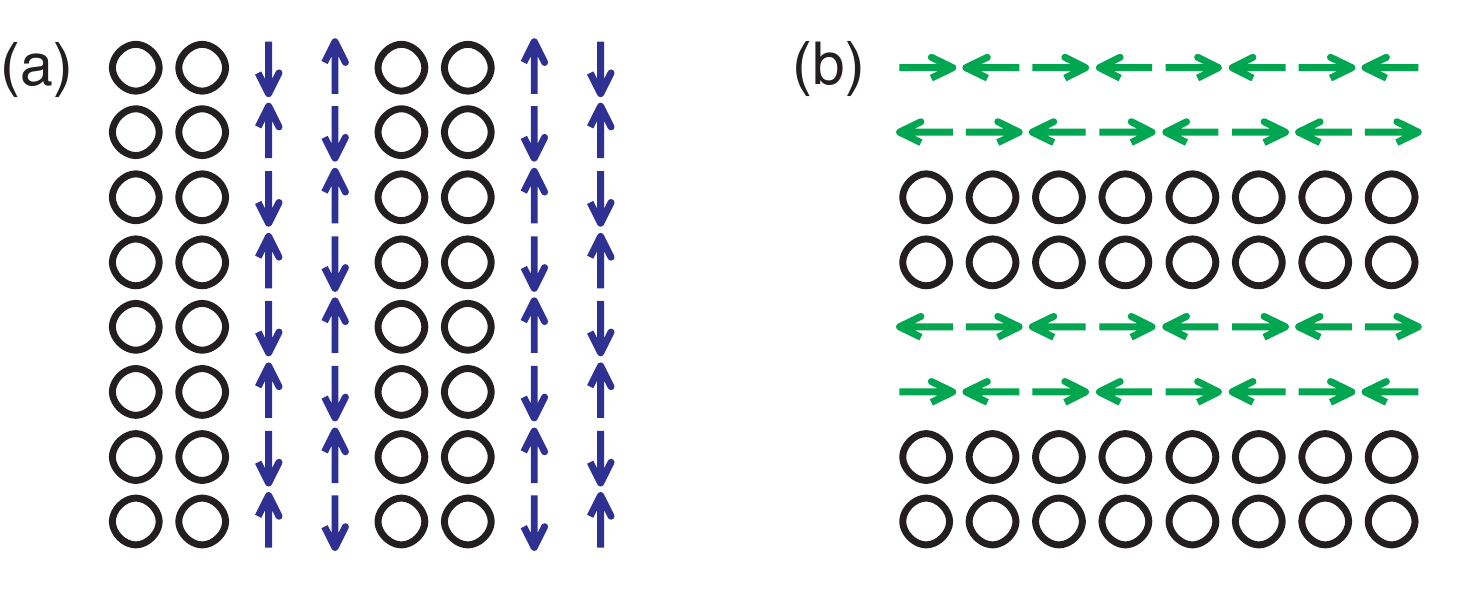}
\caption[]{ Cartoons of equivalent domains of (a) vertical and (b) horizontal bond-centered stripe order within a CuO$_2$ plane (only Cu sites shown). Note that the magnetic period is twice that of the charge period. The charge density along a stripe is one hole for every two sites in length.  The registry of the stripes with respect to the lattice (for example, site-centered vs.\ bond-centered) has not yet been determined experimentally.}
\label{Tranq:fg:stripes}\index{stripe order}
\end{figure}

Insight into the cause of the magnetic incommensurability was provided by neutron diffraction measurements on a closely related material,\linebreak 
La$_{1.48}$Nd$_{0.4}$Sr$_{0.12}$CuO$_4$ \cite{Tranq:tran95a}.   The impact of the Nd substitution is to modify the tilt pattern of the CuO$_6$ octahedra such that the in-plane Cu-O bond directions become inequivalent \cite{Tranq:axe94}.  New superlattice peaks were observed in this low-temperature phase, with in-plane wave vectors $\vec{Q}=(\frac12\pm\epsilon,\frac12)$ and $(\frac12,\frac12\pm\epsilon)$ corresponding to spin order and $(\pm2\epsilon,0)$ and $(0,\pm2\epsilon)$ associated with modulations of atomic positions due to charge order, with $\epsilon\approx0.12$.  Such results have been confirmed in the system La$_{2-x}$Ba$_x$CuO$_4$ \cite{Tranq:fuji04,Tranq:huck11}.   Analysis of the superlattice peaks indicates that they are evidence for spin and charge stripe order \cite{Tranq:kive03,Tranq:zaan01}, as illustrated in Fig.~\ref{Tranq:fg:stripes}.  Because of the crystal symmetry, the orientation of the stripes rotates $90^\circ$ from one layer to the next.

\index{La$_{2-x}$Ba$_x$CuO$_4$}

The occurrence of maximum stripe order corresponds to a strong suppression of the bulk $T_c$ at $p\approx\frac18$ \cite{Tranq:ichi00,Tranq:huck11}, suggesting that stripe order competes with superconductivity; however, recent studies have demonstrated that 2D superconductivity can coexist with stripe order \cite{Tranq:jie12}.  It now appears that superconducting order can intertwine with stripe order \cite{Tranq:berg09b}.  Thus, understanding stripe correlations may provide valuable insights into the nature of the superconducting mechanism of cuprates.

\index{La$_{2-x}$Ba$_x$CuO$_4$, magnetic dispersion}

\begin{figure}[t]
\centering
\includegraphics*[width=.7\textwidth]{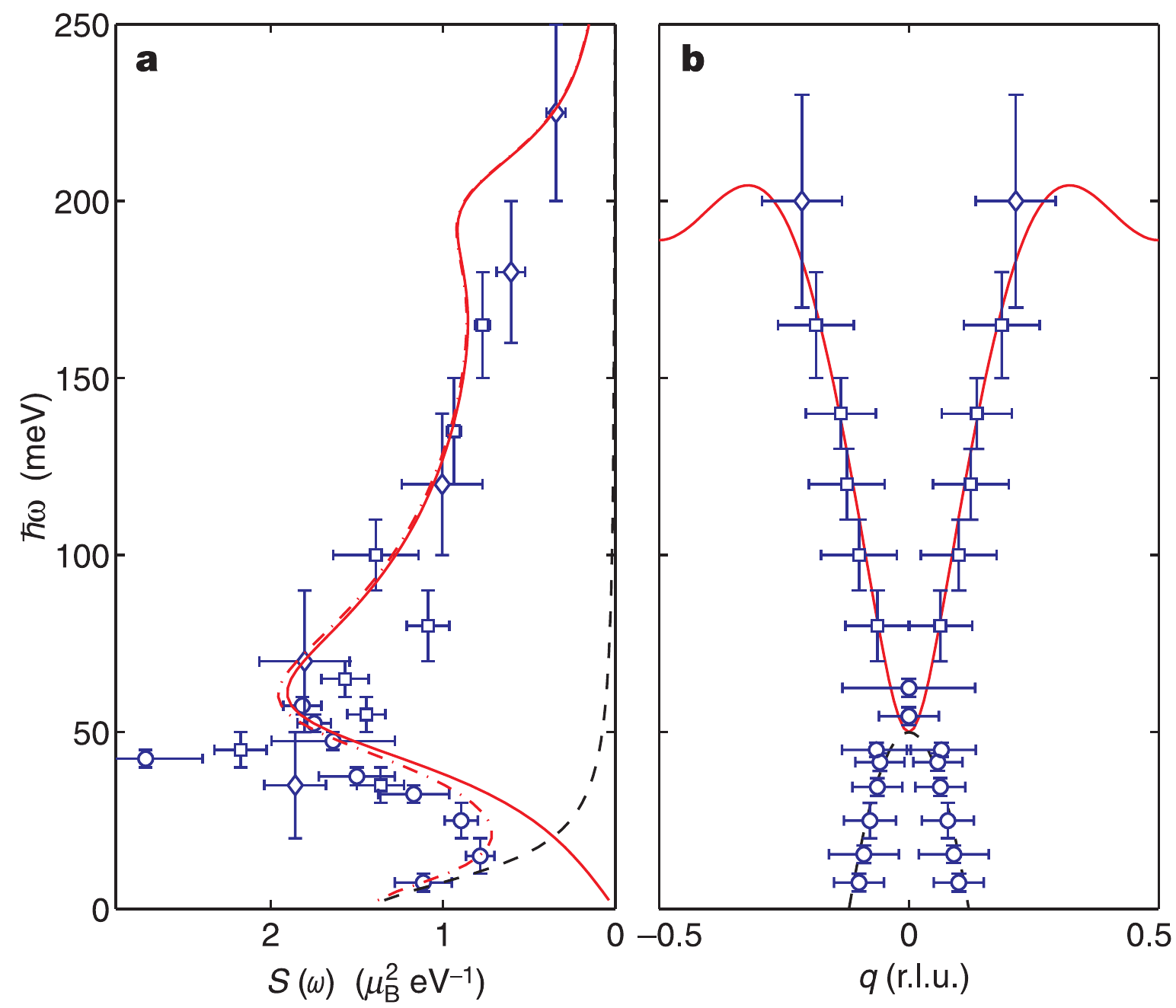}
\caption[]{(a) $\vec{Q}$-integrated spectral weight and (b) effective magnetic dispersion in the stripe-ordered phase of La$_{2-x}$Ba$_x$CuO$_4$ with $x=1/8$, from \cite{Tranq:tran04}.  The solid lines through the data points are described in the text.  In (a), the peak at $\sim40$~meV is now know to be due to a phonon mode.  In (b), the effective dispersion is plotted for $\vec{q}$ along a line through the incommensurate magnetic superlattice peaks.}
\label{Tranq:fg:disp}\index{La$_{2-x}$Ba$_x$CuO$_4$, magnetic dispersion}
\end{figure}

Neutron scattering on a time-of-flight instrument has been used to characterize the spin excitation spectrum in La$_{2-x}$Ba$_x$CuO$_4$ with $x=1/8$ \cite{Tranq:tran04}.  The effective dispersion and the $\vec{Q}$-integrated spectral weight are shown in Fig.~\ref{Tranq:fg:disp}.  Above 50~meV, the excitations disperse upwards like antiferromagnetic spin waves with an energy gap;  the solid line through the points in each panel corresponds to a two-leg spin ladder model with $J = 100$~meV.   Below 50 meV, the excitations disperse downwards toward the positions of the incommensurate magnetic superlattice peaks.  When the sample is warmed to a state with no static stripe order, the spectrum maintains its essential features \cite{Tranq:fuji04,Tranq:xu07}.  It appears that stripes, whether static or dynamic, provide a way for the superexchange mechanism to survive when the antiferromagnetic layers are doped with holes.

\index{YBa$_2$Cu$_3$O$_{6+y}$}\index{Bi$_2$Sr$_2$CaCu$_2$O$_{8+\delta}$}

The relevance of charge-stripe order is less clear in cuprates families such as YBa$_2$Cu$_3$O$_{6+y}$ and Bi$_2$Sr$_2$CaCu$_2$O$_{8+\delta}$; nevertheless, the dispersion of the magnetic excitations in these compounds (measured by neutron scattering) has been shown to be quite similar to that of LBCO \cite{Tranq:tran07,Tranq:fuji12a}.   The main difference is that the low-energy excitations tend to be gapped in the superconducting state, with a pile up of weight (``resonance'' peak) appearing above the gap for $T<T_c$.  The commonality of the dispersions over a broad energy range suggests that the charge and spin correlations in superconducting and striped cuprates are similar.


\printindex

\end{document}